\documentclass[]{aa}

\usepackage[varg]{txfonts}
\usepackage{natbib}
\usepackage[labelfont=bf]{caption}
\usepackage{adjustbox}
\usepackage[super]{nth}
\usepackage[]{siunitx}
\usepackage{subfig}
\usepackage{mathtools}
\usepackage{amsmath}
\newcommand{\uj}{ULAS J1120+0641\xspace}
\newcommand{\lya}{Ly\(\alpha\)\xspace}
\newcommand{\lyb}{Ly\(\beta\)\xspace}
\newcommand{\teff}{\(\tau_{\rm GP}^{\rm eff}\)\xspace}
\newcommand{\Hi}{{\rm H}\,\textsc{i}\xspace}
\newcommand{\xHi}{x_{\Hi}\xspace}

\usepackage{color}

\begin{document}
\urlstyle{same}
\title{Observations of the Lyman series forest towards the redshift 7.1 quasar ULAS J1120+0641} 
\author{
R. Barnett\inst{\ref{inst1}}
\and S. J. Warren\inst{\ref{inst1}}
\and G. D. Becker\inst{\ref{inst2}}
\and D. J. Mortlock\inst{\ref{inst1},\ref{inst3},\ref{inst4}}
\and P. C. Hewett\inst{\ref{inst5}}
\and R. G. McMahon\inst{\ref{inst5},\ref{inst6}}
\and C. Simpson\inst{\ref{inst7}}
\and B. P. Venemans\inst{\ref{inst8}}
}
\institute{Astrophysics Group, Blackett Laboratory, Imperial College
  London, London SW7 2AZ, UK \label{inst1}
\and Department of Physics \& Astronomy, University of California, Riverside, 900 University Avenue, Riverside, CA 92521, USA \label{inst2}
\and Department of Mathematics, Imperial College London, London SW7 2AZ, UK \label{inst3}
\and Department of Astronomy, Stockholm University, Albanova, SE-10691 Stockholm, Sweden \label{inst4}
\and Institute of Astronomy, University of Cambridge, Madingley
Road, Cambridge CB3 0HA, UK  \label{inst5}
\and Kavli Institute for Cosmology, University of Cambridge, Madingley
Road, Cambridge CB3 0HA, UK \label{inst6}
\and Gemini Observatory, Northern Operations Center, N. A'ohoku Place, Hilo HI 96720, USA \label{inst7}
\and Max Planck Institut f{\"u}r Astronomie, K{\"o}nigstuhl 17, D-69117 Heidelberg, Germany \label{inst8}
}
\date{Received <> / Accepted <>}

\abstract{We present a 30\,h integration Very Large Telescope X-shooter spectrum
  of the Lyman series forest towards the \(z = 7.084\) quasar
  \uj. The only detected transmission at ${\rm S/N}>5$ is confined to seven narrow spikes in the \lya forest, over
  the redshift range $5.858<z<6.122$, just longward of the wavelength
  of the onset of the Ly$\beta$ forest. There is also a possible detection of one
  further unresolved spike in the Ly$\beta$ forest at $z=6.854$, with
  ${\rm S/N}=4.5$. We also present revised \emph{Hubble Space
    Telescope} F814W photometry of the source. The summed flux from
  the transmission spikes is in agreement with the F814W photometry,
  so all the transmission in the Lyman series forest may have been
  detected. There is a Gunn-Peterson (GP) trough in the \lya forest
  from $z=6.122$ all the way to the quasar near zone at $z=7.04$. The
  trough, of comoving length $240\,h^{-1}$\,Mpc, is over twice as long
  as the next longest known GP trough. We combine the spectroscopic
  and photometric results to constrain the evolution of the \lya
  effective optical depth (\teff) with redshift, extending a similar
  analysis by Simpson et al. We find \teff\(\propto (1+z)^{\xi}\)
  where \(\xi = 11.2^{+0.4}_{-0.6}\), for \(z > 5.5\). The data
  nevertheless provide only a weak limit on the volume-weighted
  hydrogen intergalactic (IGM) neutral fraction at $z\sim 6.5$, \(\xHi > 10^{-4}\),
  similar to limits at redshift $z\sim6$ from less distant quasars. The new observations cannot
  extend measurements of the neutral fraction of the IGM to higher
  values because absorption in the \lya forest is already saturated
  near $z\sim6$. For higher neutral fractions, other methods such as
  measuring the red damping wing of the IGM will be required.}

\keywords{ULAS~J1120+0641}

\maketitle

\begin{figure*}[t]
\centering
\advance\leftskip-3cm
\advance\rightskip-3cm
\includegraphics[scale = 0.75, trim = 0.3cm 0.4cm 0.1cm 0.2cm, clip]{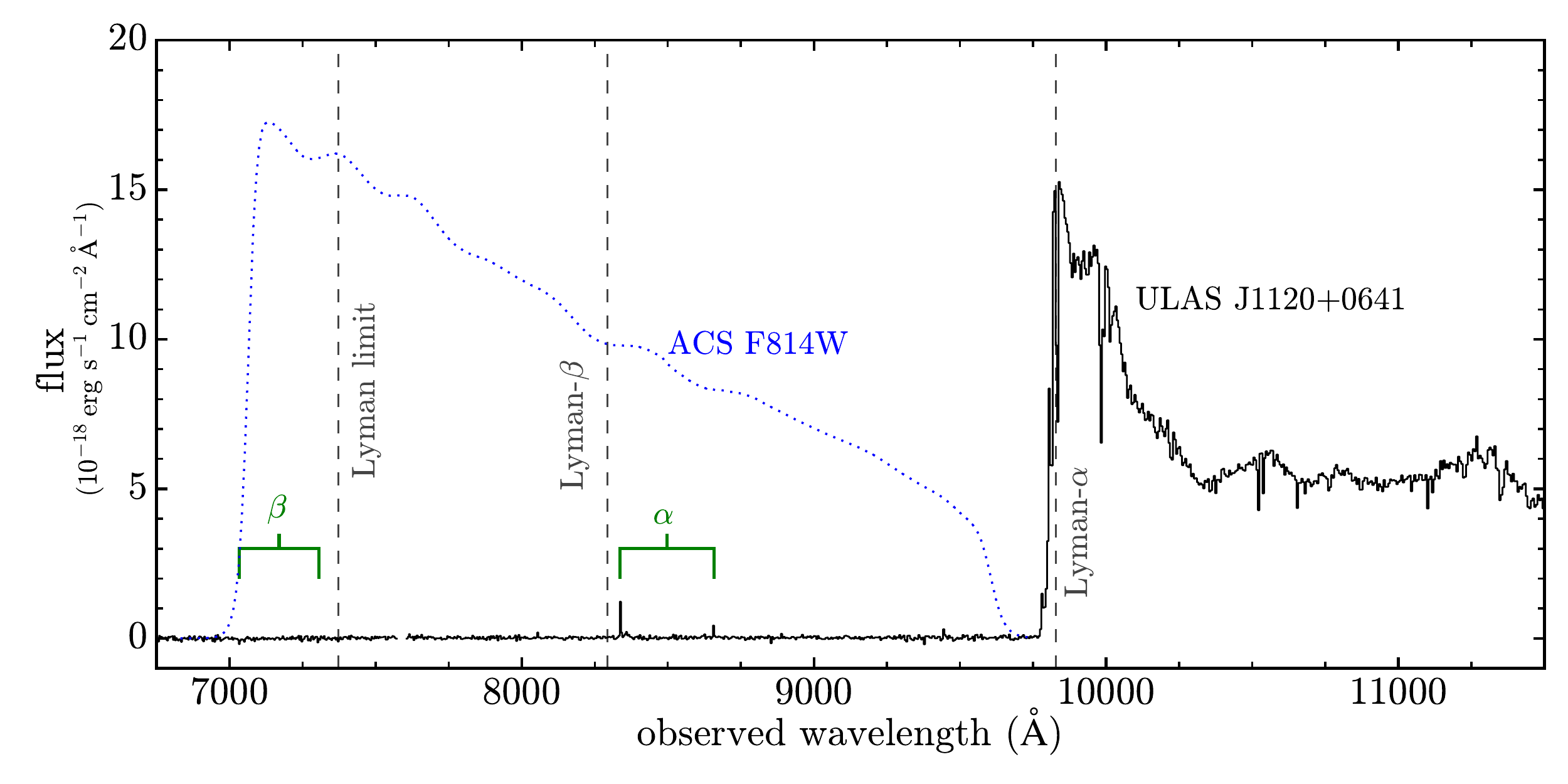}
\caption{VLT X-shooter spectrum of \uj (black), covering restframe
  wavelengths between 850 and 1400\,\AA. The spectrum has been binned to
  a pixel size of 150\,km\,s$^{-1}$ using inverse variance
  weighting. The observed wavelengths of \lya, \lyb and the
  Lyman limit are marked. The only transmission detected in the
  spectrum is confined to spikes over the wavelength range marked by
  $\alpha$, in the pure \lya forest, corresponding to the redshift range
  $5.858<z<6.122$ (Sect.~\ref{sec:spikes}). The corresponding wavelength range for Ly$\beta$
  absorption is marked $\beta$ and lies blueward of the restframe
  Lyman limit. Plotted as the blue dotted line is the (scaled) total system
  throughput curve for the ACS F814W filter.}
\label{obs}
\end{figure*}
%HST

%INTRO
\section{Introduction}
\label{sec:intro}

%The formation of the first emitting sources in the Universe brought an
%end to the so-called cosmological dark ages and triggered the epoch of
%reionization. Much observational effort has gone into understanding
%the evolution of the intergalactic medium (IGM) during this
%reionization epoch, which has provided us with insights into the
%nature of the earliest structures in the Universe. 
It is now over 50
years since \citet[GP;][]{Gunn1965} noted the lack of
Lyman-\(\alpha\) (\lya) absorption in the spectra of (at the time)
high redshift quasars. The cross-sections of the Lyman series
transitions are so high that even a small neutral hydrogen fraction (\(\xHi\sim\!10^{-4}\)) 
should result in an intergalactic medium (IGM) that is effectively opaque to photons at
wavelengths shorter than that of \lya emission,
1215.67\,\AA. However, using observations of the \(z = 2.01\)
quasar 3C 9 \citep{Schmidt1965}, \citet{Gunn1965} were able to
demonstrate an extremely low density of neutral hydrogen \((\xHi <
10^{-5}\)) in the IGM near $z=2$.

More recently, the \lya forest has been used to constrain the end of
reionization by attempting to measure the extent of \lya absorption at
redshifts of around six.
%The first \(z \sim 6\) quasars detected by the Sloan Digital Sky
%Survey \citep[e.g.][]{Fan2001} provided the first tentative evidence
%of a reionization epoch.
The lack of GP troughs in the spectra of SDSS quasars with \(z < 6\)
\citep[e.g.][]{Fan2000,Fan2001,Djorgovski2001,Songaila2002} indicated
that the IGM was highly ionized, since a low
volume weighted neutral fraction (\(\sim\!10^{-4}\)) would be
sufficient to result in complete absorption of \lya photons. The first
GP trough was reported by \cite{Becker2001} towards the \(z = 6.28\)
quasar SDSS J1010+0524 \citep[see also][]{White2003}. They concluded
that the neutral fraction must increase sharply at \(z \sim 6\).

A seminal study of Lyman series absorption was that of
\citet[][see also \citealt{Fan2006b}]{Fan2006}, who presented the largest individual sample of high
redshift quasars to date. From the spectra of 19 \(z \sim 6\) sources,
they noted a sharp increase in the effective optical depth to \lya
photons (\teff). Their conclusion was that there is an acceleration of the evolution of the
IGM ionization state at \(z \gtrsim 5.7\). Importantly,
\citet{Fan2006} reported a large line-of-sight dependence, indicating
patchiness in the IGM neutral fraction that could be related to
cosmic reionization. \citet{Fan2006} also
measured a dramatic increase in the average length of dark gaps (in
this case defined as contiguous regions with optical depth
\(\tau>3.5\)) for \(z > 6\), suggesting that \lya troughs could be a
powerful probe of the IGM at higher redshifts. 

The detection of a \(\sim\!110\,h^{-1}\)\,Mpc GP trough
covering the redshift range \(5.523 \leq z \leq 5.879\) in the \lya
forest towards the \(z = 5.98\) quasar ULAS J0148+0600 by
\citet{Becker2015} recently strengthened the case for
patchiness. Comparing the observations to hydrodynamical simulations,
they argue that significant fluctuations in the hydrogen neutral
fraction as late as \(z \simeq\) 5.6 - 5.8 are required to explain the data .

Measurements of dark pixels in the \lya and \lyb forests have also
been used to place the first model independent constraints on
reionization. Rather than attempting to measure optical depths based
on a continuum model, \citet{Mesinger2010} proposed a novel approach,
where the fraction of pixels that have zero measured flux in both
\lya and \lyb is used to directly measure an upper limit to the volume-weighted mean
neutral hydrogen fraction. Applying the idea to observed spectra,
\citet{McGreer2011,McGreer2014} conclude that reionization must be (almost)
complete by \(z \sim 6\). 

The most recent optical depth measurements from \emph{Planck}, integrated to the last scattering surface (\(z\sim1100\)), yield \(\tau = 0.055 \pm 0.009\), suggesting a model-dependent redshift of reionization between \(\sim7.8\) and \(\sim8.8\)
(\citeauthor{Planck2016a,Planck2016b} 2016).
%The highest redshift quasars may potentially reside in an IGM with a significant neutral fraction.
%\textbf{\citet{Mortlock2011} analysed \lya damping wing absorption towards the quasar, finding a lower limit on the volume-weighted neutral fraction \(\xHi > 0.1\) at \(z = 7.1\). More recently, \citet{Greig2016} measured \(\xHi = 0.4 \pm 0.2 \) at \(z =7.1\). } 

These various results both from spectroscopic analyses and \emph{Planck} motivate extending observations of \lya absorption to higher redshifts. 
\uj \citep{Mortlock2011} remains the only known quasar with \(z > 7\)
(\(z = 7.084\); \citealt{Venemans2012}). In this paper we analyse the \lya forest towards \uj by combining \emph{Hubble
Space Telescope} (\emph{HST}) imaging from the Advanced Camera for
Surveys (ACS), first presented by \citet{Simpson2014}, with new
spectroscopy from the Very Large Telescope (VLT)/X-shooter instrument
\citep{Vernet2011}. The spectrum of \uj is shown in Fig.~\ref{obs}. In
Sect.~\ref{sec:data} we present the data, paying close
attention to the effects of charge transfer efficiency (CTE) losses in
the case of \emph{HST}. We remeasure the ACS F814W ($i$-band) image of
the quasar. In Sect.~\ref{sec:specanalysis} we analyse the X-shooter spectrum. We discuss the limitations of
spectroscopy for determining optical depth measurements towards the source, and search for significant transmission in the spectrum. In Sect.~\ref{sec:disc}
we combine our photometric and spectroscopic measurements in
order to constrain the evolution of effective optical depth towards
the quasar, and use simulations to set limits on \(\xHi\).  We summarise in Sect.~\ref{sec:sum}.

All magnitudes are on the
AB system \citep{Oke1983}. We have adopted a concordance cosmology
with \(h\) = 0.7, \(\Omega_{\rm M}\) =
0.3, and \(\Omega_{\Lambda}\) = 0.7.

\section{Observations and data reduction}
\label{sec:data}

%XSHOOTER
\subsection{VLT X-shooter spectroscopy}
\label{sec:xs}

The source was observed with the X-shooter spectrograph
over several nights spanning 2011 March to 2014 April, providing a
total of 30\,h integration on source\footnote{ESO programmes 286.A-5025(A), 089.A-0814(A), and 093.A- 0707(A)}.
The Visual-red (VIS) arm
of X-shooter, with MIT/LL CCD detectors, spans the wavelength range
5500\,-\,10000\,\AA, and the Near-IR arm (NIR), with a Hawaii 2RG array detector,
spans the wavelength range 10000\,-\,25000\,\AA, or 1.0\,-\,2.5\,$\mu$m.  The quasar's \lya
emission line lies at 9827\,\AA, close to the boundary between the two
arms. In the
current paper we analyse the Lyman series absorption in the spectrum,
i.e.,  the wavelength range from the quasar Lyman limit at
7371\,\AA~up to the \lya emission line. This wavelength range lies
entirely within the VIS arm.

The data reduction, including flux calibration, is described more fully in
\citet{Bosman2}, who provide the spectrum redward of \lya, and
present the results of a search for metal lines in that region. Briefly, sky subtraction was performed using a custom {\sc idl} implementation of the optimum sky subtraction routines outlined in \citet{Kelson2003}.  This approach involves transforming the $(x,y)$ CCD coordinates into a wavelength and slit position for each pixel in the two-dimensional frame.  Slit positions for each order were defined using a standard star trace.  In orders where the quasar has little flux, such as in the \lya\ forest, the offset from the standard trace and the spatial profile of the object were extrapolated from orders where flux is present, assuming that there is minimal atmospheric dispersion between orders\footnote{The majority of our data were acquired in 2013 and 2014, when the X-Shooter atmospheric dispersion corrector was not in use.}. This assumption is reasonable given that the \lya\ forest for \uj falls at fairly red wavelengths, and because the object was generally observed at moderate airmass (less than 1.5). 
Optimal techniques \citep{Horne1986} were used to extract a one-dimensional spectrum simultaneously from all exposures for a given spectrograph arm after a relative flux calibration was applied to each order in the two-dimensional frames.

Great care was taken to minimise bias in the sky-subtraction stage of
the data reduction, since we are interested in measuring the degree of
absorption in the Lyman series forest relative to the
continuum. Systematics, where the sky level to subtract is
consistently over- or under-estimated, are more important for \uj than
for several of the other high-redshift quasars used for measuring
transmission in the \lya forest because the source is fainter. For
example, three quasars in which long GP troughs have been detected,
ULAS J0148+0600, $z=5.98$ \citep[][]{Becker2015}, and SDSS J1030+0524,
$z=6.31$ and SDSS J1148+5251, $z=6.42$ \citep{White2003}, are
respectively factors 3.9, 2.1, and 2.9 brighter than \uj.  Subtle
systematics can be present depending on the algorithm used. For
example, a weighted least squares fit to the sky where the variance is
a Poisson estimate based on the measured counts will systematically
underestimate the sky and so leave a positive residual (see
\citealt{White2003} for a related discussion). To avoid such a bias,
we instead used a least-squares linear fit.  Likewise, 
combining multiple exposures using inverse variance weighting can produce a small positive bias when the variance is calculated using a Poisson estimate based on amplitude of the sky fit (plus a contribution from read noise). The bias occurs because exposures in which the sky is underestimated (and hence the object flux overestimated) at a particular wavelength will be assigned a lower variance, and hence a larger weight.  To mitigate this bias, we estimated the variance as a function of wavelength using the measured scatter in the two-dimensional counts about the sky fit. A Poisson contribution from the object, based on the extracted counts, is also added along the object trace.

%separate exposures were combined using inverse variance weights, for which we
%used the measured scatter about the fit rather than a Poisson estimate
%from the counts as an estimate of the sky variance.

The observations were made with a 0\farcs9 slit width, providing a
nominal resolving power of \(R = 8800\), or a resolution of
34\,km\,s$^{-1}$. Inspection of the telluric absorption lines,
however, indicates that the true mean resolving power was somewhat
better, \(R \sim 10000\), or a resolution of 30\,km\,s$^{-1}$, as a
consequence of good average seeing during the observations. The bin
size used for the final spectrum is 10\,km\,s$^{-1}$. The spectrum was
flux calibrated using observations of a spectrophotometric
standard. Due to slit losses the zero point (i.e. overall scaling)
will be uncertain. Accordingly the spectrum was then scaled by matching to the
GNIRS spectrum presented in \citet{Mortlock2011} over the region
of overlap, redward of the \lya forest, 9800\,-\,10000\,\AA.  The
GNIRS spectrum itself was calibrated to UKIRT photometry, from
2011. The quasar was observed in 2013 March with {\emph HST}
(Sect.~\ref{sec:hst}) in the F814W, F105W, and F125W filters
\citep{Simpson2014}, and found to have faded by 0.16 mag. compared to
the UKIRT \(YJ\) photometry. Because we use the F814W photometry in
our analysis of the \lya forest, we have applied this additional flux
correction to the spectrum, meaning that the normalisation differs
from that used by \citet{Bosman2}.

The new X-shooter spectrum is presented in Fig. \ref{obs}, binned to a pixel size of 150\,km\,s$^{-1}$. As
detailed below, the only detectable transmission over the entire
Lyman series is confined to the small range in wavelengths marked by
$\alpha$ in Fig. \ref{obs}, corresponding to the redshift range
5.858\,-\,6.122, in the pure \lya forest, and just redward of the onset
of the Ly$\beta$ forest, at $\lambda_{\rm obs} = 8291.92\,\AA, z=5.821$. The remainder
of the \lya forest from $z=6.122$ up to the edge of the quasar near zone
at $z=7.04$ is completely absorbed, as is the entire region of the Ly$\beta, \gamma$ and higher series, down
to the Lyman limit. The bin width of 150 km\,s$^{-1}$ in Fig. \ref{obs}
was chosen to emphasise the GP absorption in the plot.

We checked the computed error spectrum by analysing the distribution
of signal-to-noise (S/N) ratio in pixels in the long GP trough. We
first subtracted off any large-scale systematic residual
(Sect.~\ref{sec:system}) using a 301 pixel median filter, before
calculating the S/N ratios. We applied a \(3\sigma\) clip and fit a
Gaussian to the resulting histogram, which should have a dispersion of
unity if the errors are correct. We bin the data for this particular
analysis, e.g., by factors of 2, 4, etc., to account for the slight
smoothing at the pixel-to-pixel level introduced by rebinning at the
wavelength calibration stage. Scaling the noise array by a factor of
1.05 resulted in a Gaussian fit to the residuals with a dispersion of
unity. As noted by \citet{Bosman2}, the sky subtraction is imperfect
close to the peak of the brightest OH sky lines. Again fitting a
Gaussian to the S/N histogram for pixels close to the peak of strong
lines, we were able to treat the imperfect sky subtraction by further
increasing the errors by a factor of 1.15 for the five pixels centred
on each skyline peak.

%The spectrum may be the deepest yet taken of a quasar with $z>5.8$. 
% GDB: There are other z~6 QSOs that have X-Shooter spectra with higher S/N (see Becker+2015).

The errors are on average substantially smaller than in the spectra of the three quasars listed earlier. For example compared to \uj, in the GP troughs in the spectra of ULAS J0148+0600, $z=5.98$, SDSS J1030+0524, $z=6.31$, and SDSS J1148+5251, $z=6.42$, measured over the same wavelength regions, the errors are on average larger by factors 1.6, 2.5, and 1.8, respectively (consistent with the difference in integration times).

\begin{table*}[t]
\centering
\advance\leftskip-3cm
\advance\rightskip-3cm
\caption{Measured flux density in GP troughs. The final column represents the flux measured over the same wavelength range for \uj, with significant transmission spikes subtracted.}
\label{tab:troughs}
\begin{tabular}{ccccc}
\hline \hline
quasar & redshift range & wavelength range & flux density & \uj  
\\[0.1ex]
     &    &  \AA &   10$^{-20}\)\,erg\,s$^{-1}$\,cm$^{-2}$\,\AA$^{-1}$  & 10$^{-20}\)\,erg\,s$^{-1}$\,cm$^{-2}$\,\AA$^{-1}$ \\
\hline \\[-2ex]
ULAS J0148+0600 & $5.55-5.71$ & $7966-8163$ & $\!\!\!\!-0.3\pm0.9$ & $2.2\pm0.5$ \\
ULAS J0148+0600 & $5.71-5.88$ & $8163-8360$ & $0.9\pm0.8$ & $2.7\pm0.5$ \\
SDSS J1030+0524 & $6.00-6.16$ & $8510-8710$ & $1.5\pm1.4$ & $3.2\pm0.5$ \\
SDSS J1148+5251 & $6.10-6.32$ & $8630-8900$ & $3.5\pm0.8$ & $2.7\pm0.5$ \\
\hline
\end{tabular}
\end{table*}

\subsection{\emph{HST} imaging}
\label{sec:hst}

\begin{figure}[t]
\centering
\subfloat[]{
        \label{subfig:cte1}
        \includegraphics[width=0.47\textwidth]{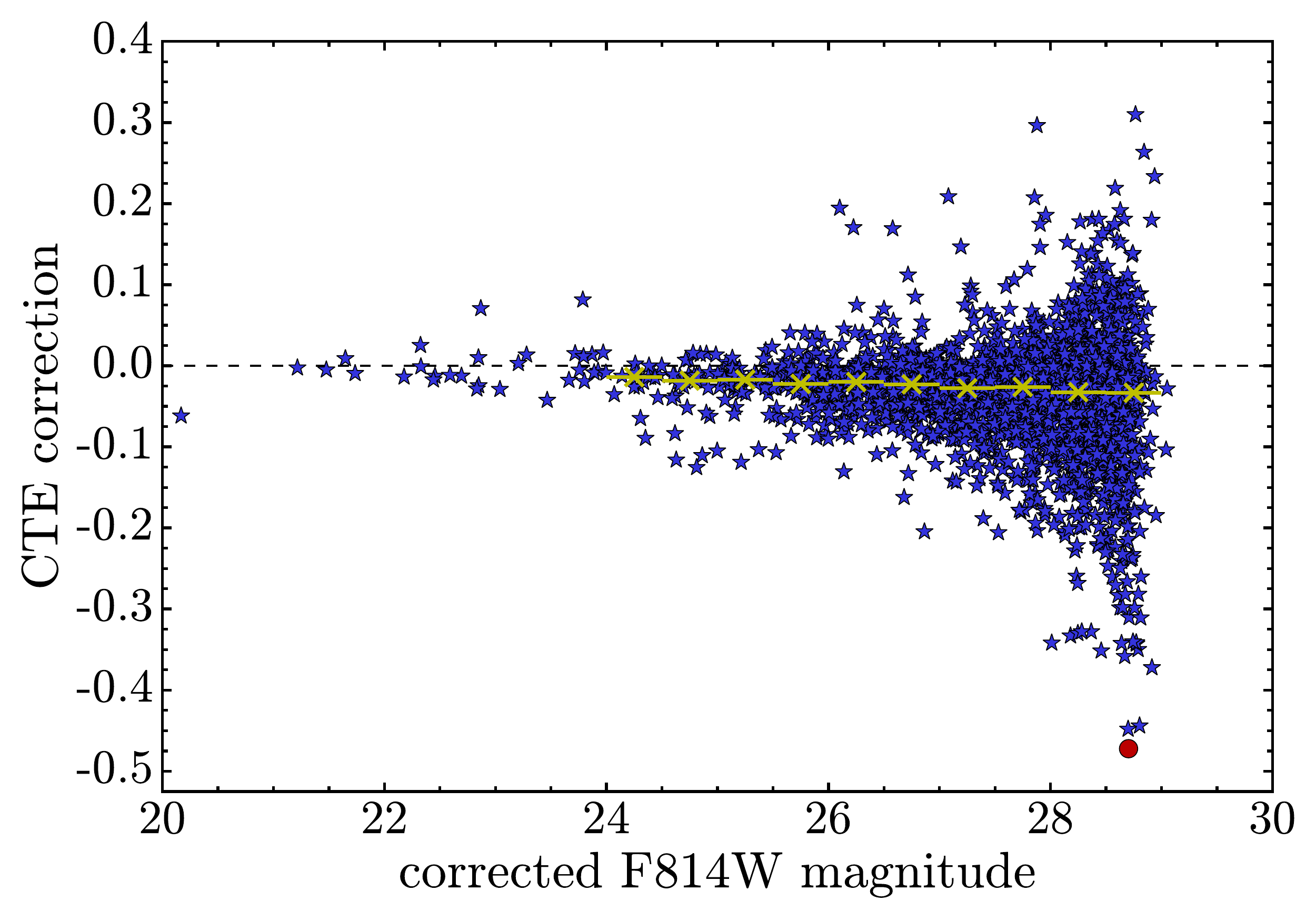} }

\subfloat[]{
        \label{subfig:cte2}
        \includegraphics[width=0.47\textwidth]{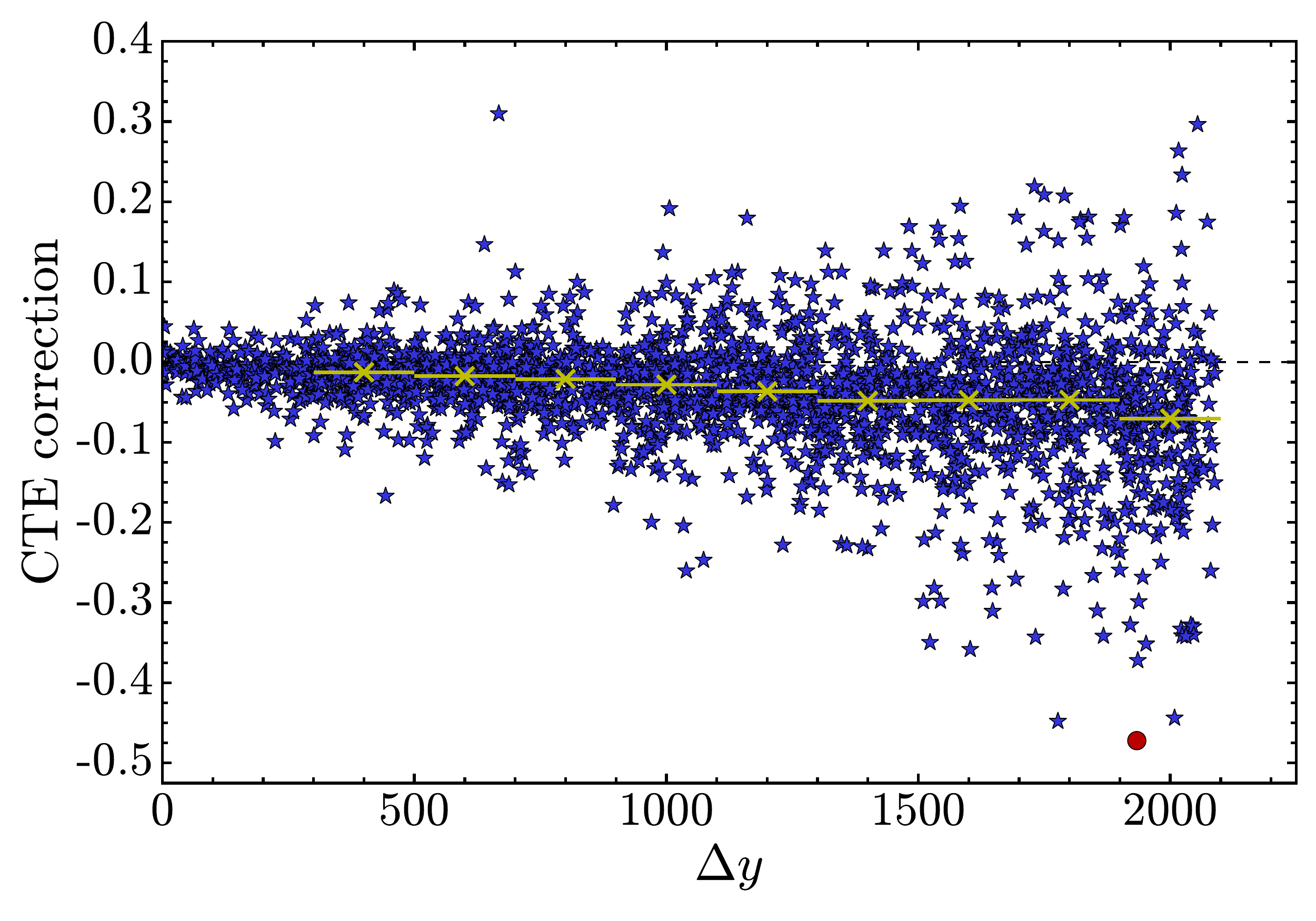} } 
\caption{The CTE correction (\(m_{\rm corrected} - m_{\rm uncorrected})\)
  applied to objects detected in the F814W frame, shown as
  blue stars. The red circle indicates \uj. Yellow crosses represent the median correction in the indicated bins of 0.5 mag (upper) or 200 pixels (lower). \textbf{Upper:} Correction as a function of corrected F814W magnitude. \textbf{Lower:} Correction as a function of \(\Delta y\), the number of pixels from the readout register.}
\label{ctecorrs}
\end{figure}

\emph{HST} observed the field of \uj with the ACS and WFC3
instruments in the three filters
F814W, F105W, and F125W, during 2013 March \citep{Simpson2014}.
%The goal of the observations was to identify Lyman-break galaxies associated with the quasar.
As shown in Fig. \ref{obs} the wavelength range of
the ACS F814W filter rather neatly spans the Lyman series forest of
\uj, and so photometry of \uj in this filter provides valuable
information on the Lyman series absorption. The F814W image had a
total exposure time of 28,448s spread across five visits made over 2013
March 27--29; it is substantially deeper than the Subaru
$i^\prime$-band image presented in \citet{Barnett2015}. Each visit
comprised six dithered exposures, with the exception of the
final visit where a single exposure was taken. The 25
pipeline-processed frames were then combined using the \textsc{pyraf}
routine \emph{astrodrizzle}.

\citet{Simpson2014} measured \(i_{\rm814} = 28.59 \pm
0.20\) for \uj with aperture photometry, using the result to constrain
the evolution of the effective optical depth towards the quasar
(Sect.~\ref{sec:tau}). The source is close to the limit of detectability,
and in ACS images the effect of charge transfer efficiency (CTE)
losses can be important for such faint sources. As the counts in a
detected source are clocked down to the serial register, some charge
gets left behind and appears in a trail after the source
\citep{Anderson2010,Chiaberge2012}. The problem is more severe the
lower the background and proportionally greatest for the faintest
sources, increasing linearly with distance of the source from the
serial register, i.e., with the parallel clocking distance. The
placement of \uj close to the centre of the field of view of the ACS
instrument unfortunately means that the object lies at the edge of one
of the CCDs. The distance to the serial register is maximised, as is the charge loss. Therefore, it is important to assess the
effect of CTE losses on the photometry.

Two solutions to the problem of CTE losses have been devised by the
ACS team. The first \citep{Chiaberge2012} quantifies the losses as a
function of sky background, source brightness, and location on the
CCD, and provides formulae to correct measured photometry. The second
\citep{Anderson2010} attempts to put the trailed charge back into the
sources, by applying an ingenious algorithm that uses a model for the CCD
traps (defects) that cause the CTE losses. The latter method has
proved so successful that it is now applied by default in the ACS data
reduction pipeline (although there is an option to turn it off), and
the image measured by \citet{Simpson2014} was corrected in this
manner.

\uj is located close to the ecliptic, resulting in a comparatively
bright sky background. In a single frame the average sky level in our
F814W frames is \(\sim\)105\(e^-\). This is a considerably higher
background than the levels explored by \citet{Chiaberge2012}. As the
background level rises the CTE corrections become smaller, and the
coefficients provided by \citet{Chiaberge2012} allow one to estimate
the background level at which the CTE correction becomes
zero. Although it becomes necessary to extrapolate their results, the
level is $\sim 70 e^-$,
%RB: I have checked this by plotting the alpha and beta parameters (calculated using Table 1 data) from Chiaberge report against sky level and 70e seems to be about right (i.e. the paramters ----> 0), although value gets lower for later cycles.
 implying that any CTE corrections in our frames should actually be very small. We
consulted with the ACS instrument team on the issue of the CTE
correction, and they recommended that
at such high background levels losses should indeed be small, and
that we should use frames without the \citet{Anderson2010} corrections
applied, since for high background the algorithm will simply add noise.

As a check, we used \emph{astrodrizzle} to produce co-added frames
of the dithered F814W exposures both with and without the
\citet{Anderson2010} correction in order to quantify the CTE
corrections made as a function of source brightness, and of location on the CCD. 
We paid close attention to the choice of input parameters
to \emph{astrodrizzle} to avoid the cores of bright objects being
masked as cosmic rays.
%We used \textsc{SExtractor} \citep{Bertin1996} to
%produce a catalogue of objects in the \emph{HST} frames, limited to
%point sources, since the CTE losses will be different for extended
%sources. The centroid of the image of \uj is uncertain as the source is so faint. Therefore, %the catalogue was defined using the
%F105W image. We performed aperture photometry at the positions
%specified by the F105W catalogue on the corrected and uncorrected
%F814W images. Finally, we calculated the CTE corrections \((\Delta\,m =
%m_{\rm corrected} - m_{\rm uncorrected})\) for all sources.
We used \textsc{SExtractor} \citep{Bertin1996} to
produce a catalogue of objects in the \emph{HST} frames,
detected in the corrected F814W image. We performed aperture photometry
at the positions specified by the catalogue in both the corrected and
uncorrected F814W images. Ideally the sample would be restricted to
point sources, but star/galaxy discrimination is unreliable for
sources as faint as \uj, so we included all catalogued sources. CTE
corrections for galaxies will be larger than for point sources, as
they cover more pixels, so any measured trends may in fact
overestimate the CTE correction for a point source. 
The centroid of the image of \uj is also uncertain as
the source is so faint. We therefore used the F105W frame to define the position
of the quasar, before measuring photometry in the two F814W frames. However, the F105W 
proved a poor choice to define the whole catalogue, as it is shallower than
the F814W observation and so fewer objects were detected.
Finally, we calculated the CTE corrections \((\Delta\,m = m_{\rm corrected} - m_{\rm uncorrected})\) for all sources.
These are presented as a function of corrected magnitude in
Fig.~\ref{subfig:cte1}, and of the distance in pixels from the serial
register, \(\Delta y\), in Fig.~\ref{subfig:cte2}.
%If a CTE correction
%were required we would expect to see trends in the size of the CTE
%correction with both magnitude and \(\Delta y\). Instead, the mean correction appears to be
%consistent with zero, and no trends are apparent.
%Fig.~\ref{ctecorrs} suggests that when the background is this high
%the \citet{Anderson2010} algorithm simply introduces additional noise,
%and confirms that the uncorrected frame provides the best photometry.
%The correction for \uj itself is anomalous: in general \(\Delta m\) is
%quite small, \(\sim 0.1\) mag., except in the case of the quasar,
%which is brightened by \(0.47\) mag. We do not have a satisfactory
%explanation for the discrepancy. Nevertheless, there are very few
%sources fainter than 28th magnitude, so the scatter at
%such faint magnitudes is not well established, and may be larger than that
%suggested by the plot.
In each of these plots very weak trends, at the level of few hundredths of a
magnitude are visible in the mean correction for all objects at a
particular magnitude or clocking distance, but these are much
smaller than the standard deviation. Furthermore, as noted above, these trends may
overestimate the required correction, as the sample is dominated by
galaxies. We conclude that when the background is this high
the \citet{Anderson2010} algorithm simply introduces additional noise,
and the best photometry is provided by the uncorrected frame.
The correction for \uj itself is large: the quasar lies in the tail of
the distributions and is brightened by
0.47\,mag. This unusually large correction appears to be due to chance.

The uncorrected frame has a stripy appearance from (fractionally
small) CTE losses in bright sources. Therefore, we flattened the sky
background before aperture photometry by subtracting a median-filtered
version of the frame. We measured the counts in a 3-pixel radius
(rather than the 2.5 pixel aperture preferred by
\citealt{Simpson2014}, to minimise digitisation issues), and applied
an aperture correction to 10 pixel radius ($0\farcs5$), derived from bright
stars. This was then further corrected to the nominal ``infinite''
aperture radius of the ACS of $5\farcs5$ using the quoted encircled
energy fraction of 0.914 in the $0\farcs5$ aperture, quoted in the
ACS Data Handbook.
The final photometry result is
\(i_{\rm814} = 28.85\), with \({\rm S/N} = 2.9\). We quote the error
as a signal-to-noise ratio as the errors are Gaussian in flux, but at
such low significance the equivalent posterior distribution in
magnitudes is extremely skewed, and we do not quote it here.

\begin{figure*}[t]
\centering
\advance\leftskip-3cm
\advance\rightskip-3cm
\includegraphics[scale = 0.9, trim = 2cm 1.5cm 0.8cm 0.2cm, clip]{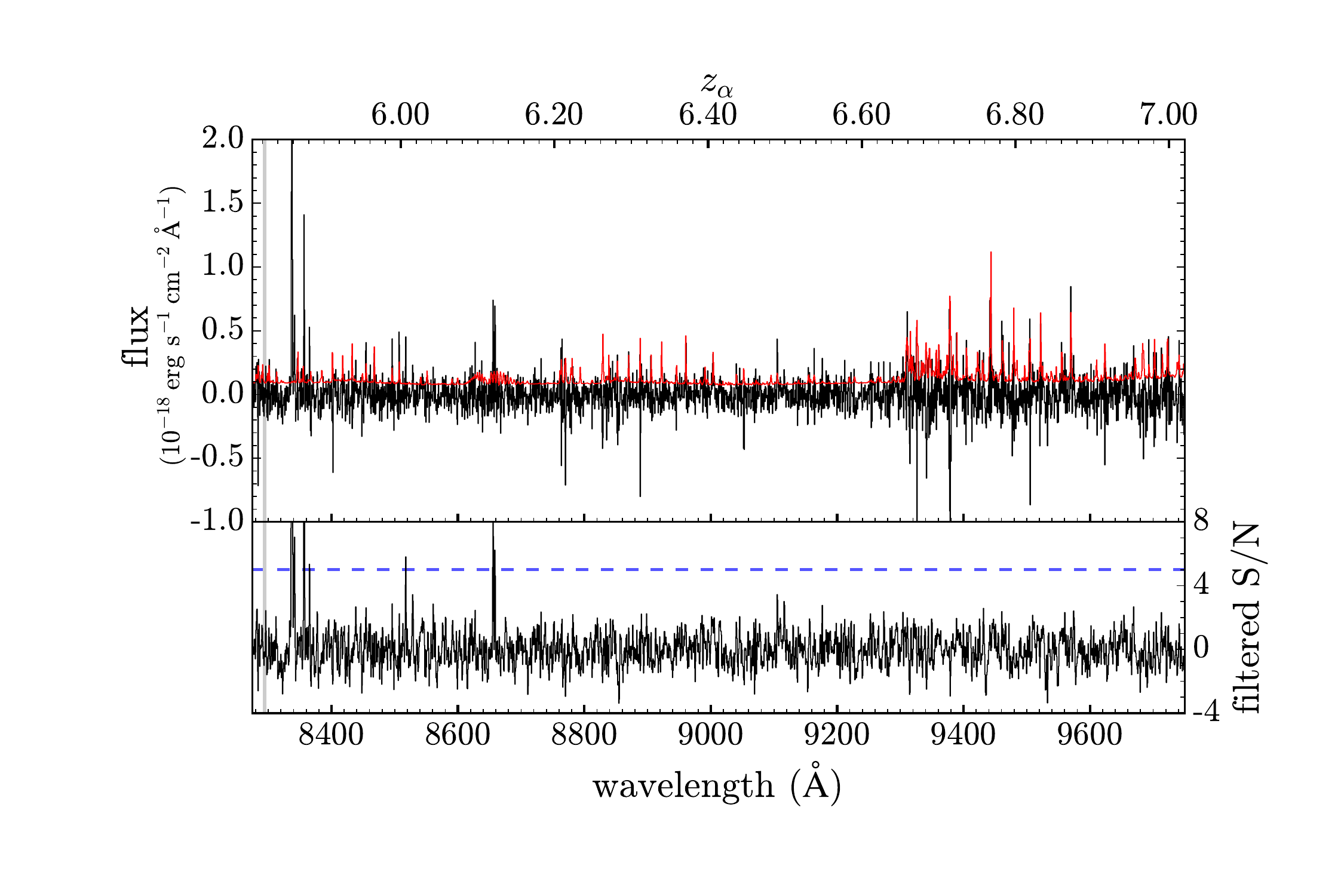}
\caption{Close-up of the \lya forest towards \uj, binned to a pixel size of 20\,km\,s$^{-1}$ for display purposes. \textbf{Upper:} Spectrum (black) and \(1\sigma\) errors (red). \textbf{Lower:} S/N at each pixel found by running the Gaussian matched filter at the narrowest width over the unbinned spectrum (pixel size 10\,km\,s$^{-1}$).}
\label{lya}
\end{figure*}

\section{Analysis of the spectrum}
\label{sec:specanalysis}

\begin{figure}[t]
\centering
\advance\leftskip-3cm
\advance\rightskip-3cm
\includegraphics[scale = 0.35, trim = 0cm 0.5cm 0cm 0.0cm, clip]{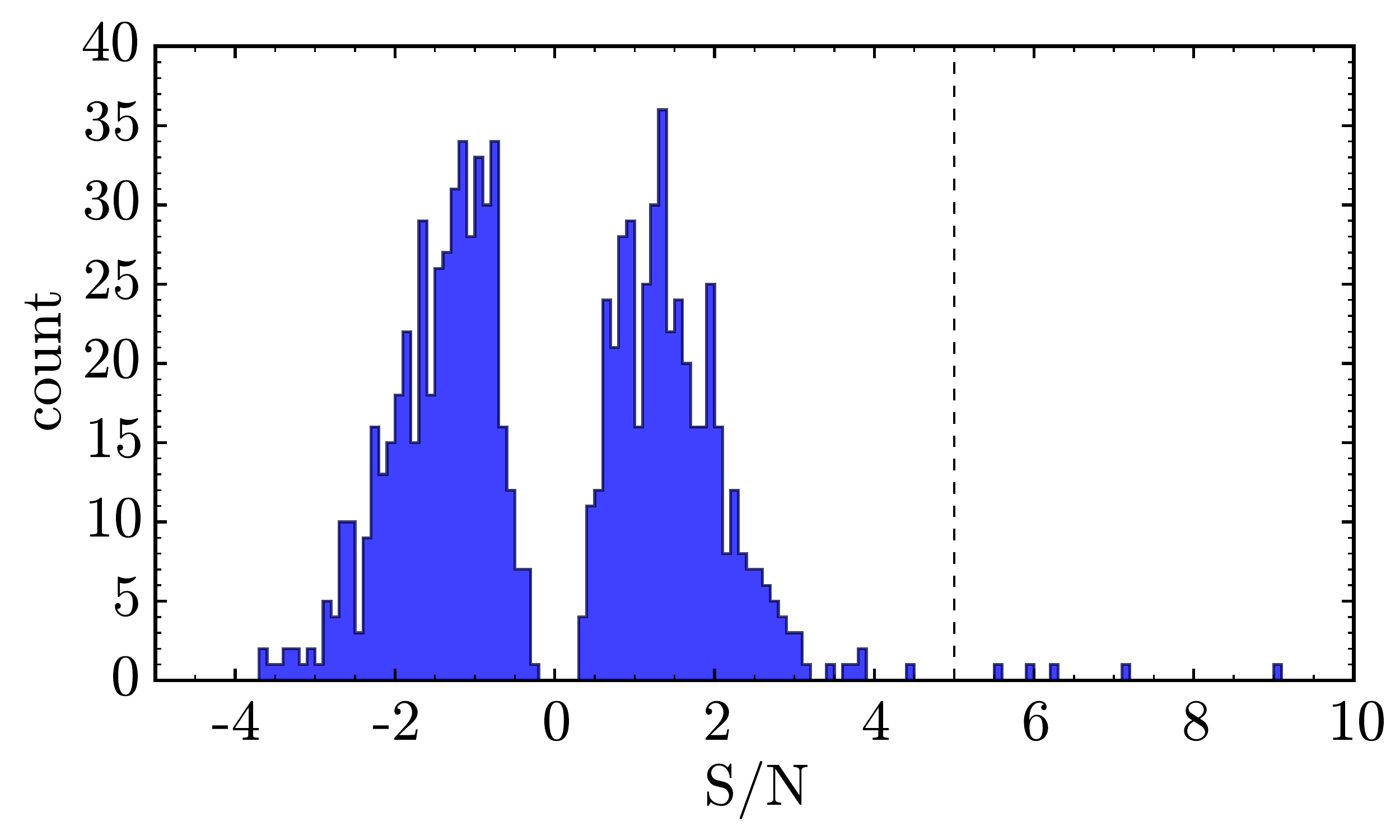}
\caption{S/N of spikes extracted from the X-shooter spectrum. We have extracted the highest value of |S/N| from all template widths.}
\label{allsnrs}
\end{figure}

\begin{figure*}[t]
\centering
\advance\leftskip-3cm
\advance\rightskip-3cm
\includegraphics[scale = 0.6, trim = 0cm 0.5cm 0cm 0.0cm, clip]{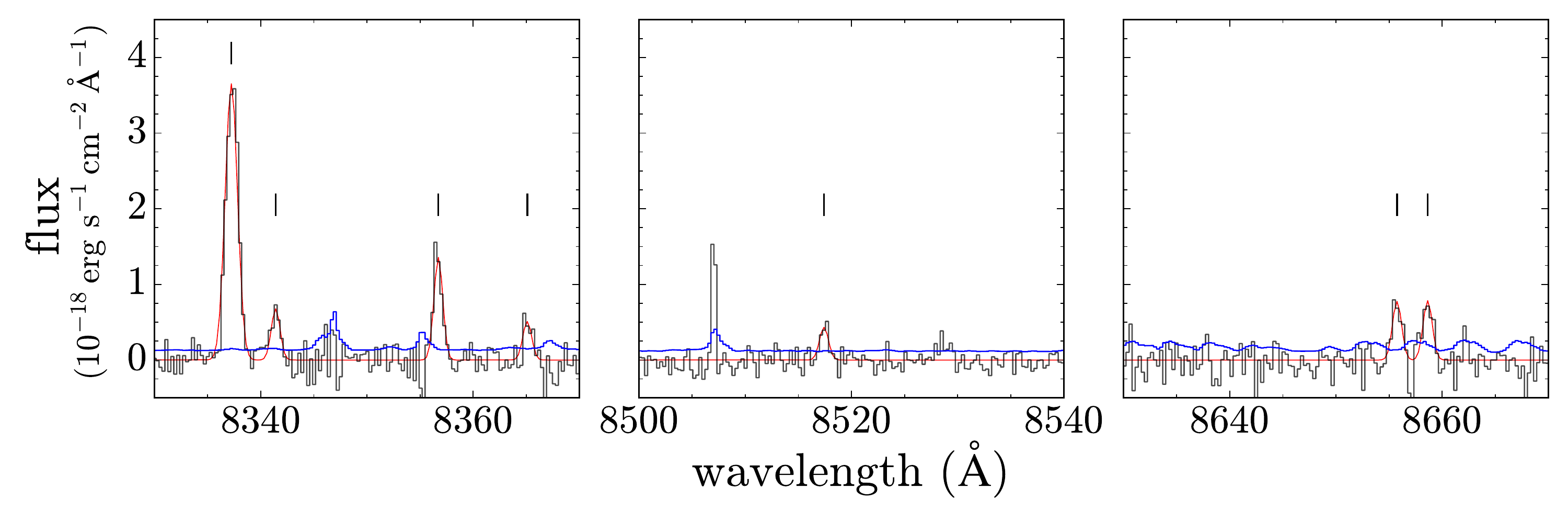}
\caption{Significant transmission of flux in the \lya forest towards
  \uj. The unbinned X-shooter spectrum of \uj is shown in grey with a pixel
  size of 10\,km\,s\(^{-1}\). Shown in blue is the 1\(\sigma\) error spectrum,
  and the derived transmission spectrum is overlaid in red. The seven
  detections presented in Table~\ref{tab:spikes} are indicated.}
\label{stamps}
\end{figure*} 

\subsection{Systematic errors in sky subtraction}
\label{sec:system}

In exploring the \lya forest in the new spectrum, smoothing on
different scales, we found that despite the care taken in the data
reduction to minimise such effects (Sect.~\ref{sec:xs}), there are
systematic errors present in the form of a small positive residual
signal at an average level of
$f_\lambda\sim3\times10^{-20}$\,erg\,s$^{-1}$\,cm$^{-2}$\AA$^{-1}$. The
excess corresponds to only \(\sim0.05\%\) of the counts from the sky
within the extraction aperture, or \(\sim0.15\) of the average random
error per pixel. We used \textsc{synphot} \citep{Laidler2008} to
measure synthetic photometry from the spectrum, obtaining a value
\(i_{\rm814} = 27.2\), which is a factor 4.5 brighter than the
\emph{HST} photometry. The low-level sky residual is difficult to
characterise quantitatively, since a \(3\sigma\) detection alone
requires binning over 400 pixels ($\sim120$\,\AA). However, the
residual does not correlate directly with the error spectrum, in the
sense that on large scales they do not share the same shape. Neither
is the residual significantly greater in the wavelength ranges of the
bands of OH sky lines, compared to the line-free regions.  Therefore
it does not seem to be an effect associated with the difficulty of
subtracting the sharp OH sky lines, and is apparently a smooth
feature.

Systematic errors in sky subtraction have not been reported or
quantified in previous measurements of the Lyman-series effective
optical depth.
%At a level similar to that reported here they would have been detectable only in very deep observations of long GP troughs. To quantify this statement 
We checked whether a residual at the level seen in the spectrum of \uj
would have been detected in the GP troughs seen in the spectra of the
three quasars ULAS J0148+0600 \citep{Becker2015}, SDSS J1030+0524, and
SDSS J1148+5251 \citep{White2003}. In these papers, the transmission
in the GP troughs is measured in bins of width $\sim50$\,Mpc, and
transformed to an effective optical depth. Using the same bins we have
measured the average flux density in the spectra, using
inverse-variance weighting, and the results are provided in the fourth
column of Table \ref{tab:troughs}. In three of the bins the measured
flux is consistent with zero, while significant flux is recorded in
the GP trough $6.10<z<6.32$ in the spectrum of SDSS J1148+5251. We
then measured the average flux density over the same four bins in the
spectrum of \uj, after subtracting the few transmission spikes seen in
the spectrum (Sect.~\ref{sec:spikes}), in order to measure the smooth
residual. These results are listed in the final column of Table
\ref{tab:troughs}. In the spectrum of \uj we measure significant flux
at the level of
$\sim3\times10^{-20}$\,erg\,s$^{-1}$\,cm$^{-2}$\AA$^{-1}$ in all four
bins.

The detection of significant flux in the redshift bin $6.10<z<6.32$ in
the spectrum of SDSS J1148+5251 is interesting, because it is the only
other case where significant flux has been measured where no
transmission spikes in \lya have been detected. In the original paper,
\citet{White2003} speculated that this was due to flux from a faint
intervening galaxy, a suggestion which has subsequently been ruled out
by \emph{HST} imaging \citep{White2005}. Systematic errors in the sky
subtraction may provide a simpler explanation. In fact, comparing the
values in columns 4 and 5 of Table \ref{tab:troughs}, in each bin the
measurements agree at the $2.5\sigma$ level. This means that
systematic errors could be present in the other spectra at a similar
level, and the reason they are reported here for the first time is
because it is the deepest spectrum yet taken at these redshifts.

\begin{table}[t]
\centering
\advance\leftskip-3cm
\advance\rightskip-3cm
\caption{Transmission spikes detected in the \lya forest with the matched filter.}
\label{tab:spikes}
\begin{tabular}{ccccc}
\hline \hline
\(\lambda_{\rm obs}\) & \(z_{\alpha}\) & \(\Sigma\) & S/N & Spike flux 
\\[0.1ex]
    \AA &    &   km\,/\,s &   & 10\(^{-18}\)\,erg\,s\(^{-1}\)\,cm\(^{-2}\) \\
\hline \\[-2ex]
8337.2  & 5.858 & 21 & 48.7 & 5.34 \\
8341.4  & 5.862 & 15 & 7.4  & 0.71 \\
8356.7  & 5.874 & 15 & 14.8 & 1.42 \\
8365.1  & 5.881 & 15 & 5.6  & 0.54 \\
8517.4  & 6.006 & 15 & 5.9  & 0.46 \\
8655.7  & 6.120 & 15 & 9.0  & 0.84 \\
8658.6  & 6.122 & 15 & 6.2  & 0.85  \\
\hline
\end{tabular}
\end{table}

\subsection{Detection of transmission spikes}
\label{sec:spikes}

To remove the systematic residual we subtracted a median-filtered
version of the spectrum using a box of length 301 pixels, i.e.,
$\sim90$\,\AA.  At these redshifts transmission in the Lyman-series
forest will be in isolated spikes, so the application of a median
filter of such a long box length should have a negligible effect on
any inferred transmission.  The region of the spectrum covering the
\lya forest, binned by a factor two for display purposes, i.e., pixels
of 20\,km\,s$^{-1}$, is shown in Fig. \ref{lya}, together with the
error spectrum. Note, however, that we use the unbinned spectrum to
search for regions of significant transmission.  To search for spikes
in the Lyman series forest (\(7370\,\AA \leq \lambda_{\rm obs} \leq
9770\,\AA\)) we employ a Gaussian matched filter
\citep{Hewett1985,Bolton2004}, with a range of widths. The narrowest
profile has a standard deviation $\Sigma=15$\,km\,s$^{-1}$,
corresponding to the nominal spectral resolution of 34\,km\,s$^{-1}$
FWHM. A range of profiles successively increasing in width by
$\sqrt{2}$ was employed, $\Sigma=15$, 21, 30, \ldots,
120\,km\,s$^{-1}$. We fit a Gaussian at every pixel, 8535 in
  total. To illustrate the process we plot the S/N at each pixel for
  the narrowest template in the bottom panel of Fig. \ref{lya}. Then,
  for each pixel, we recorded the \({\rm S/N}\) of the most significant
  detection over all the templates, whether positive or negative, and
  the corresponding value of \(\Sigma\).

The distribution of these S/N values is shown in
  Fig.~\ref{allsnrs}. This distribution is non-Gaussian, because we
  recorded the maximum absolute value for each pixel. Since negative
spikes, which are not real, extend down to nearly ${\rm S/N}=-4$, we
cannot be sure positive spikes are real unless the S/N is
substantially greater than 4. In fact one might expect a somewhat more
extended tail in the positive direction, of spikes that are not real,
because of cosmic rays that are not perfectly subtracted. For these
reasons we set our detection threshold at ${\rm S/N}=5$.  The matched
filter finds seven significant flux transmission spikes, at observed
wavelengths \(8337\,\AA \leq \lambda_{\rm obs} \leq 8659\,\AA\),
redshifts $5.858<z<6.122$.  These are summarised in
Table~\ref{tab:spikes} and plotted in Fig.~\ref{stamps}. These seven
spikes are also identifiable in both panels of Fig. \ref{lya}.  To
confirm that the detections are real, for each spike we compared the
measured flux at the detected wavelength in subsets of the data, and
found no significant differences between subsets. We also checked the
goodness of the fit of the Gaussian template to the data, finding the
\(\chi^2\) values to be satisfactory. The observed transmission
features are very narrow: with the exception of the largest spike at
8337\,\AA, the template width \(\Sigma\) which yielded the highest S/N
was 15\,km\,s\(^{-1}\), i.e., the spikes are unresolved. We integrate
the template transmission spectrum to find a total transmitted flux of
(\(1.02 \pm 0.03\)) \(\times 10^{-17}\)\,erg\,s\(^{-1}\)\,cm\(^{-2}\),
of which approximately half is contained in the largest spike.

Using \textsc{synphot} we measure \(i_{\rm814} = 29.00\pm0.03\) for these seven spikes. Our value is consistent with the \emph{HST} photometry (Sect.~\ref{sec:hst}; \(i_{814} = 28.85\), ${\rm S/N} = 2.9$), raising the question of whether the seven spikes account for all the transmission in the spectrum.
In considering this question we refer to Table \ref{abc} which sets out the wavelength ranges in the spectrum and the corresponding redshifts over which absorption in the four transitions \lya, $\beta$, $\gamma$, $\delta$ are recorded. It is noticeable that the seven spikes all lie close to the blue end of the pure \lya forest, which is to be expected. Starting at the red end of the \lya forest, on the edge of the quasar near zone (near 9770\,\AA), and moving to shorter wavelengths, the \lya opacity reduces, down to 8291\,\AA, $z_\alpha=5.82$, where the high-opacity Ly$\beta$ forest at $z_\beta=7.08$ cuts in. Moving to shorter wavelengths from 8291\,\AA\, the opacity in Ly$\beta$ becomes lower, but will still be very high at 7862\,\AA, corresponding to $z_\beta=6.61$, at which point the Ly$\gamma$ forest cuts in at $z_\gamma=7.08$. The result is that the most transparent window in the spectrum is just redward of 8291\,\AA. Therefore if there is any more transmitted flux in the spectrum, which will be in the form of narrow spikes below $5\sigma$ significance, most of it is expected to lie in the wavelength range 8291\,-\,8700\,\AA. Looking at the spikes with S/N in the range 3\,-\,5 these are approximately uniformly distributed along the spectrum, with no excess detected in the wavelength range 8291\,-\,8700\,\AA. Even if all the 3\,-\,5\,$\sigma$ spikes in this wavelength range were real, they would only contribute an additional $10\%$ to the transmitted flux. 
%Checked by RB - 10% is right. There are only 2 spikes at 3-5sigma over this wavelength range and the distribution is pretty uniform over the whole spectrum - nothing to suggest they are real
The region of Ly$\beta$ absorption corresponding to the transmission spikes, is marked $\beta$ in Fig. \ref{obs}. It is beyond the quasar Lyman limit, and is consequently not useful for measuring transmission.

\begin{figure}[t]
\centering
\advance\leftskip-3cm
\advance\rightskip-3cm
\includegraphics[scale = 0.6, trim = 0cm 0.5cm 0cm 0.0cm, clip]{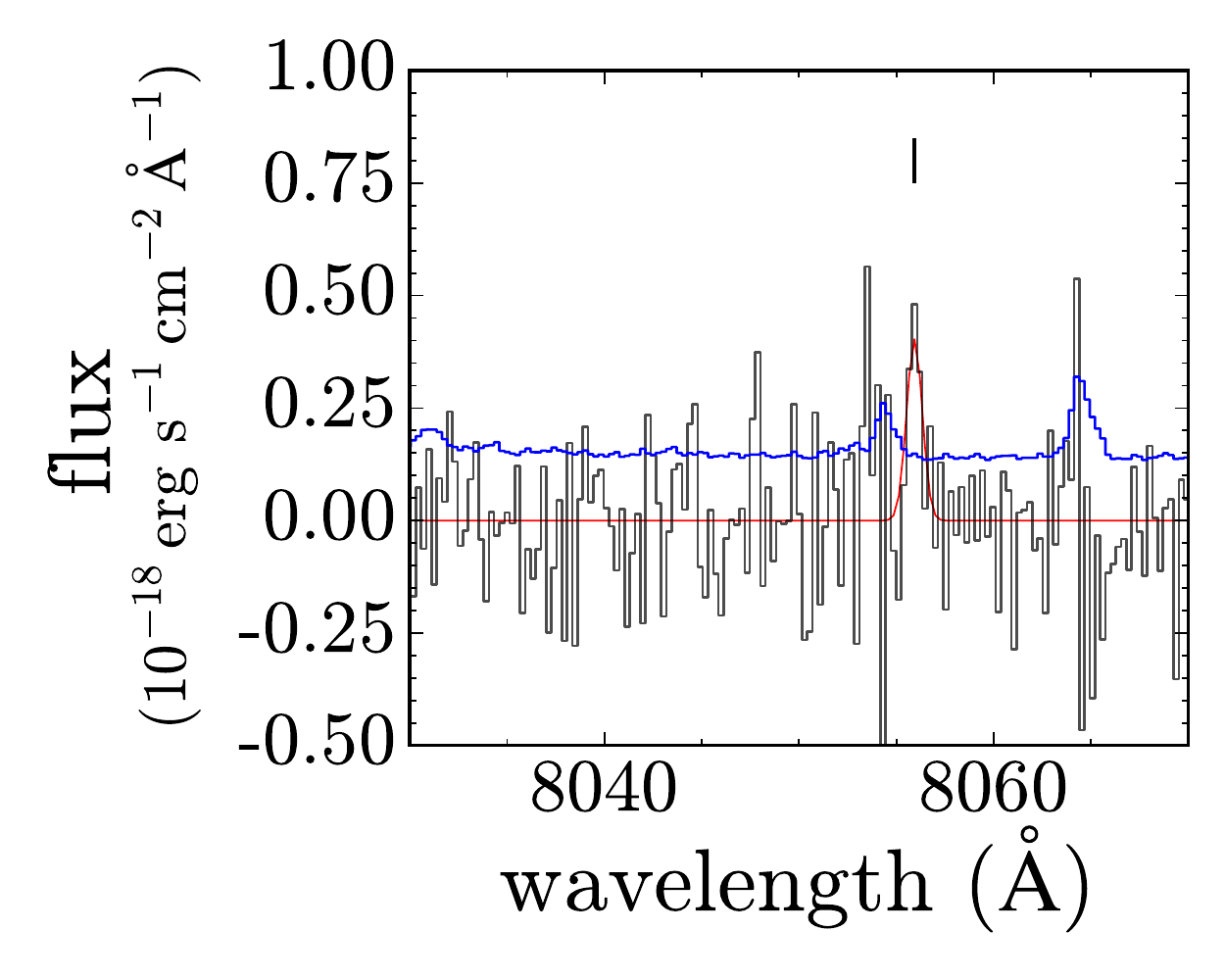}
\caption{As Fig.~\ref{stamps}, but focused on the region of the spectrum in the Ly$\beta$ forest in the vicinity of the possible transmission spike at \(\lambda = 8056\,\AA\), which is indicated.}
\label{lybspike}
\end{figure}

\begin{table}[t]
\centering
\scriptsize
\advance\leftskip-3cm
\advance\rightskip-3cm
\caption{Wavelength and redshift ranges recording \lya, $\beta$, $\gamma$, $\delta$ absorption.}
\label{abc}
\begin{tabular}{ccccc}
\hline \hline
\(\lambda_{\rm obs}\) & $\alpha$ & $\beta$ & $\gamma$ & $\delta$ \\[0.5ex]
    \AA &    &   & & \\
\hline \\[-2ex]
$8291-9827$  & $5.82-7.08$ & $-$ & $-$ & $-$ \\
$7862-8291$  & $5.47-5.82$ & $6.61-7.08$ & $-$ & $-$ \\
$7678-7862$  & $5.32-5.47$ & $6.49-6.66$ & $6.89-7.08$ & $-$ \\
$7581-7678$  & $5.24-5.32$ & $6.39-6.49$ & $6.80-6.89$ & $6.98-7.08$ \\
\hline
\end{tabular}
\end{table}

There is a single unresolved spike in the range of ambiguity
$4<{\rm S/N}<5$. It is detected at ${\rm S/N}=4.5$, and lies at \(\lambda =
8056\,\AA\), corresponding to $z_\alpha=5.627$ and
$z_\beta=6.854$. The spike is plotted in Fig. \ref{lybspike}.  A
region of transmission at $z=6.854$ would be very unexpected given the
rapid evolution of the hydrogen neutral fraction seen around $z\sim
6$, so the feature is potentially interesting. We checked that the
fluxes measured in subsets of the data are consistent. The putative
line lies six pixels from the centre of a strong OH sky line,
sufficiently far away that it should not be affected by poor sky
subtraction. Unfortunately it is not possible to be sure if the line
is real or not.

If the spike is real, it requires that the opacities along the line of
sight to the quasar at redshifts $z_\alpha=5.627$ and $z_\beta=6.854$
are both low. We are able to quantify this statement by following a
line of reasoning used by \citet{White2005} in discussing a Ly$\beta$
transmission spike in the spectrum of SDSS J1148+5251. In the
spectrum of \uj, the fractional transmission measured over five pixels centred on the \lyb spike is
\(\mathcal{T}_{8056}=0.04\pm0.01\). The effective optical depth at 8056\,\AA,
which is the sum of contributions from both the \lya and Ly$\beta$
forests, is
$\tau_{8056}=3.2\pm0.3=\tau_\alpha(z=5.6)+\tau_\beta(z=6.9)$. The
transmission at the corresponding wavelength in the \lya forest,
$z_\alpha=6.854$, $\lambda=9548$\,\AA , measured over 5 pixels, is
\(\mathcal{T}\)(9548)$=-0.01\pm0.01$, meaning an optical depth
$\tau_\alpha(z=6.9)>4.0$ at 2$\sigma$. Taking the ratio of effective
optical depths $\tau_\alpha/\tau_\beta\sim2.25$ from \citet{Fan2006},
the measurement at $\lambda=9548$\,\AA, implies
$1.8<\tau_\beta(z=6.9)<3.2$, and $0<\tau_\alpha(z=5.6)<1.4$. The range
on $\tau_\alpha(z=5.6)$ is not unreasonable, meaning that our analysis
does not weaken the case that the line is real. Combining the two
transmission measurements, and assuming
$\tau_\alpha/\tau_\beta\sim2.25$, leads to the constraint
$4.0<\tau_\alpha(z=6.9)<7.2$ for this spike.

%One might speculate that the \lyb spike points to a local ionised
%region, surrounding star forming galaxies. In this case one would expect
%the region to be enriched in metals, yet there is no evidence for
%metal lines corresponding to the redshift of the spike in the spectrum
%\citep{Bosman2}.

\subsection{Measurement of the effective GP optical depth}

\begin{table*}[t]
\centering
\scriptsize
\advance\leftskip-3cm
\advance\rightskip-3cm
\caption{Measurements of the transmission fraction \(\mathcal{T}\) and effective optical depth \teff in bins of \(\Delta z = 0.15\) towards \uj. Lower limits on \teff correspond to twice the uncertainty on \(\mathcal{T}\), if the measurement of \(\mathcal{T}\) is below 2\(\sigma\).\vspace{1pt}}
\label{tab:teff}
\begin{tabular}{ccccrc}
\hline \hline
wavelength range & redshift range & \multicolumn{2}{c}{bin size} & \multicolumn{1}{c}{\(\mathcal{T}\)} & \teff \\[0.5ex]
 \AA              &  \(z_{\alpha}\) &  pixels & \(h^{-1}\)cMpc   &                         &\\
\hline \\[-2ex]
\(7381.3-7563.5\) & \(5.07-5.22\) &  732    & 53.6   & \(0.00000 \pm 0.00088\) & \(>6.34\) \\
\(7563.7-7745.8\) & \(5.22-5.37\) &  635    & 51.7   & \(-0.00114 \pm 0.00110\)& \(>6.12\) \\
\(7746.0-7928.2\) & \(5.37-5.52\) &  698    & 49.9   & \(0.00111 \pm 0.00093\) & \(>6.29\) \\
\(7928.5-8110.7\) & \(5.52-5.67\) &  682    & 48.3   & \(0.00067 \pm 0.00093\) & \(>6.29\) \\
\(8110.9-8292.9\) & \(5.67-5.82\) &  666    & 46.6   & \(0.00015 \pm 0.00082\) & \(>6.41\) \\
\(8293.1-8475.5\) & \(5.82-5.97\) &  653    & 45.2   & \(0.00640 \pm 0.00095\) & \(5.05\pm0.15\) \\
\(8475.8-8657.8\) & \(5.97-6.12\) &  638    & 43.7   & \(0.00013 \pm 0.00084\) & \(>6.38\) \\
\(8658.1-8840.2\) & \(6.12-6.27\) &  625    & 42.4   & \(-0.00023 \pm 0.00091\)& \(>6.31\) \\
\(8840.5-9022.5\) & \(6.27-6.42\) &  612    & 41.1   & \(0.00055 \pm 0.00094\) & \(>6.28\) \\
\(9022.8-9204.9\) & \(6.42-6.57\) &  600    & 39.9   & \(-0.00058 \pm 0.00086\)& \(>6.37\) \\
\(9205.2-9387.5\) & \(6.57-6.72\) &  589    & 38.8   & \(0.00071 \pm 0.00118\) & \(>6.05\) \\
\(9387.8-9569.6\) & \(6.72-6.87\) &  576    & 37.5   & \(-0.00019 \pm 0.00114\)& \(>6.09\) \\
\hline
\end{tabular}
\end{table*}

\begin{figure}[t]
\centering
\advance\leftskip-3cm
\advance\rightskip-3cm
\includegraphics[scale = 0.35, trim = 0cm 0.5cm 0cm 0.0cm, clip]{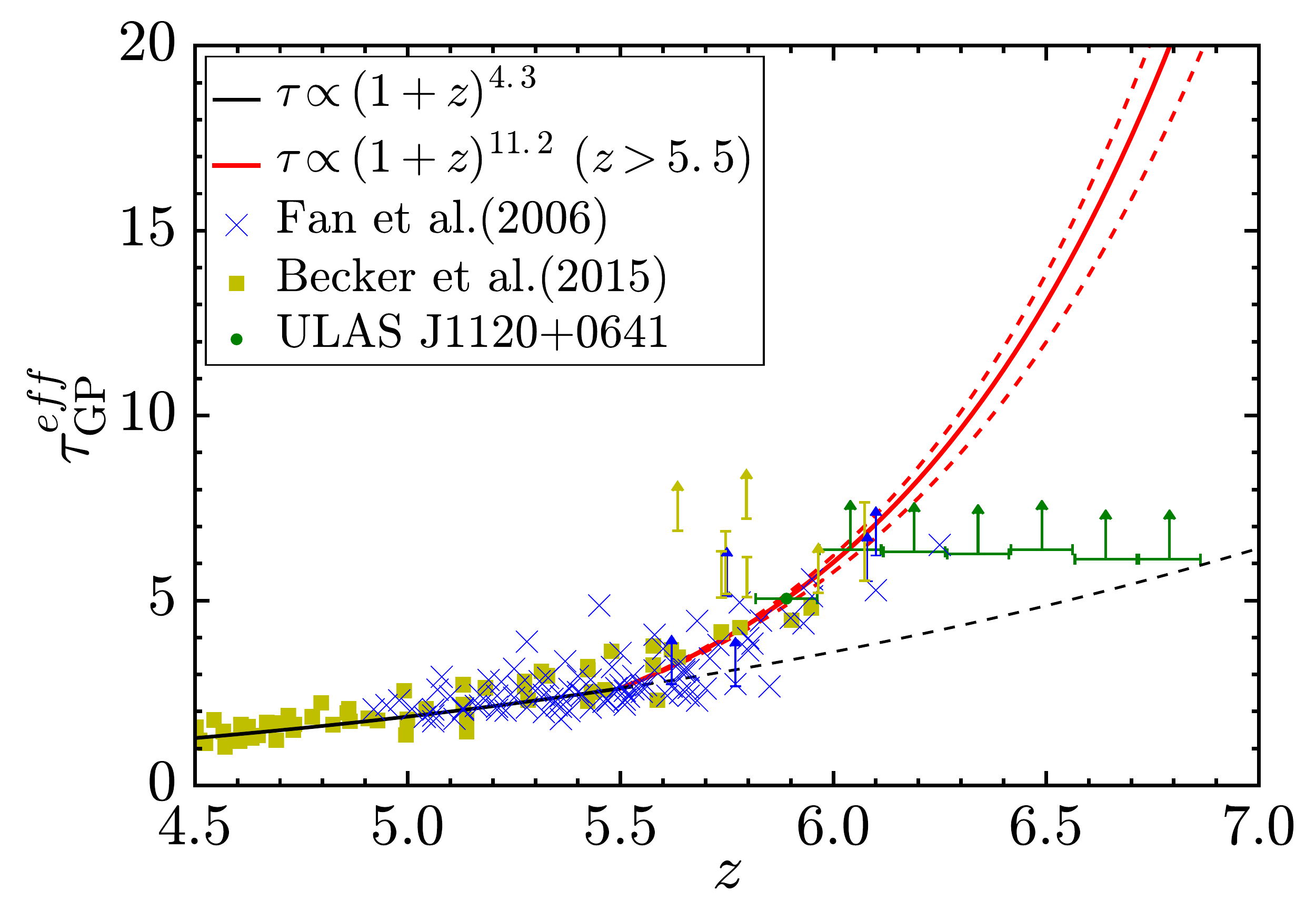}
\caption{Evolution of \teff with redshift as predicted by
  Equation~\ref{eq:tauz} (solid lines). The red curve is determined using \uj data alone and assumes a fixed normalization at \(z = 5.5\). The uncertainty in \(\xi\) for \(z > 5.5\) is indicated by the red dashed lines. The black dashed curve indicates the
  expected optical depth if the low-redshift case were to continue
  indefinitely. Points indicate direct measurements of \teff along
  different lines of sight.}
\label{tauevo}
\end{figure}

%Because the long median filter should not affect any genuine transmission spikes, we consider it reasonable to 
Narrow features are essentially unaffected by the application of the long median filter, and so we
use the filtered spectrum to measure the effective optical depth in the Lyman series
forest. To normalise the spectrum and hence estimate the transmitted
flux ratio \(\mathcal{T}\), we divide the X-shooter spectrum by the model intrinsic spectrum
described in \citet{Mortlock2011}. Here the blue wing of the \lya
emission line comes from a composite spectrum created from sources
with similar CIV blueshifts, and the spectrum is extended bluewards as
a power law \( F_\lambda\propto\lambda^{-0.5}\).

In Table~\ref{tab:teff} we present measurements of \(\mathcal{T}\) and
effective GP optical depth \teff over the redshift range \(5.07\,\textrm{-}\,7.02\),
covering practically the entire Lyman series forest. We follow the
prescription of \citet{Fan2006}, making measurements in fixed-size
bins of \(\Delta z = 0.15\). The redshift bin $6.87\,\textrm{-}\,7.02$ was excluded
as the median filtered spectrum is affected by the quasar \lya emission line in
this region. To estimate \(\mathcal{T}\) we calculate the transmission
fraction in each pixel before taking the inverse-variance weighted
average for each bin, i.e., \(\mathcal{T} = \left \langle {F/F_{\rm cont}} \right \rangle \). If we detect transmission below a level
of \(2\sigma\), we adopt a lower limit for \teff based on a value of
\(\mathcal{T}\) equal to twice the uncertainty\footnote{
A more elegant approach would be to deduce a posterior probability density function for \(\mathcal{T}\) and hence \teff. The true value of \(\mathcal{T} \geq 0\), so by assuming a uniform prior for \(\mathcal{T} \geq 0\) and a Gaussian likelihood, it would be possible to determine posterior confidence intervals for \teff.
 By adopting a \(2\sigma\) upper limit for \(\mathcal{T}\), we have followed the method used to measure existing limits on \teff \citep[e.g.][]{Fan2006,Becker2015}. Given that absorption in the \lya forest is saturated at \(z \sim 6.5\), our choice is ultimately not very important.
 }.

We also consider the effect of the adopted power-law slope at wavelengths below \lya. We recalculate \teff values using a bluer continuum, \( F_\lambda\propto\lambda^{-1.4}\) \citep{Lusso2015}. The effect on \teff is found to be small. In the lowest-redshift bin we find \teff \(> 6.54\), while in the highest redshift bins the difference is negligible. In the bin beginning at \(z = 5.82\) we measure \teff \(= 5.15 \pm 0.15\), i.e., consistent with the value presented in Table~\ref{tab:teff}.

In Fig.~\ref{tauevo} we show our measurements of \teff, as well as the
power law evolution of \teff determined using \emph{HST} photometry
(Sect.~\ref{sec:tau}). We do not include the bins for which \(z_{\alpha} <
5.82\) in Fig.~\ref{tauevo}, since these have contributions from \lyb
and higher order transitions, as well as \lya. In principle,
observations of \lyb in the redshift range $6.61<z<7.08$ might provide
a more sensitive measurement of the effective optical depth, if
corrected for the coincident absorption by \lya at
$5.47<z<5.82$. Although a statistical correction for foreground \lya is possible at
lower redshifts, the recent results of
\citet{Becker2015} show that this is not possible at \(z\gtrsim5.5\), as some lines of sight are completely absorbed in \lya.

The results for \uj lend support to the steadily-growing body of
evidence that \teff climbs rapidly at high redshift, and that
therefore we are observing the tail-end of the epoch of
reionization. However, it is also clear from Fig.~\ref{tauevo} that we
are unable to make especially useful measurements of \teff for \(z
\gtrsim 6.5\). Absorption in the spectrum becomes saturated, and so no
further information can be gained by measuring up to redshifts of
seven since we can only determine lower limits which are similar in
value to the measurements of \teff from lower redshift sources. As expected, the most useful constraints on \teff come from \(z\sim6\).

\section{Analysis and discussion}

\label{sec:disc}

%TAU
\subsection{IGM opacity}
\label{sec:tau}

\citet{Fan2006} used the spectra of 19 \(z > 6\) quasars to measure \teff along different lines of sight. They found the best-fit power law to be:
\begin{equation}
\label{eq:tauz}
\tau_{\rm GP}^{\rm eff}(z) = 
  \begin{dcases*}  
    \,0.85 \left(\frac{1 + z}{5} \right)^{4.3} \quad  z \leq 5.5\\
    \\[-2.5ex]
    \,2.63 \left(\frac{1 + z}{6.5} \right)^{\xi}\hphantom{^{.3}}   \quad z > 5.5\\    
  \end{dcases*},
\end{equation}
empirically determining a limit of \(\xi > 10.9\). We now wish to
measure a value of \(\xi\) following a procedure similar to that used
by \citet{Simpson2014}, using our revised photometry. We have an
additional constraint on \(\xi\), namely the transmission spikes
detected in the X-shooter spectrum.

\begin{figure}[t]
\centering
\advance\leftskip-3cm
\advance\rightskip-3cm
\includegraphics[scale = 0.3, trim = 0cm 0.5cm 0cm 0.0cm, clip]{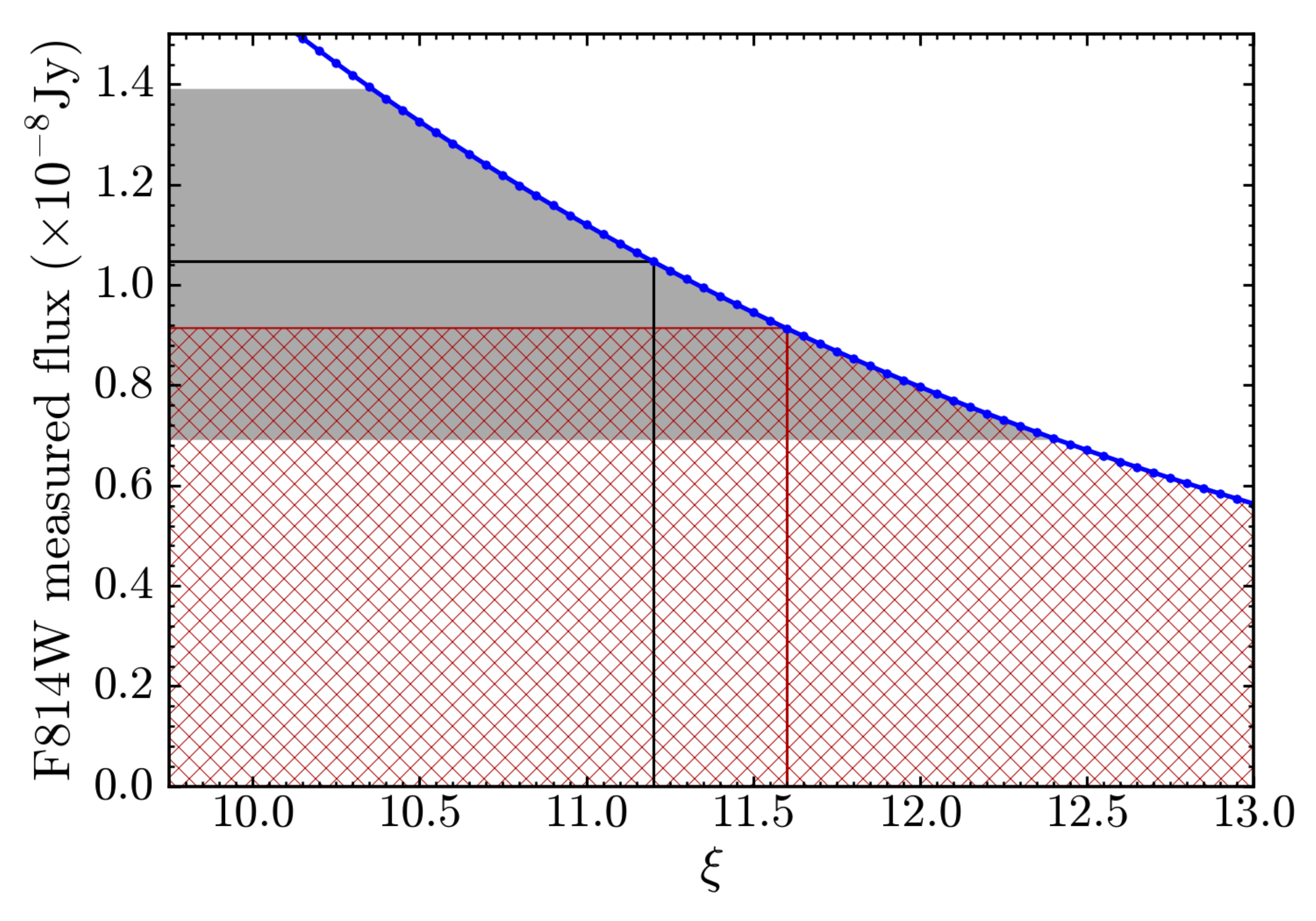}
\caption{Synthetic F814W photometry measured by determining \teff from different values of \(\xi\) (Eq.~\ref{eq:tauz}) is indicated by the blue points. The solid black line represents the F814W flux in the {\it HST} ACS frame. The \(1\sigma\) range for the {\it HST} measurement is shaded grey. The red hatching indicates the hard lower limit on the measured F814W flux imposed by the X-shooter spectrum. The vertical red line indicates the resulting upper limit on \(\xi\).}
\label{xis}
\end{figure}

\begin{figure}[t]
\centering
\advance\leftskip-3cm
\advance\rightskip-3cm
\includegraphics[scale = 0.35, trim = 0cm 0.5cm 0cm 0.0cm, clip]{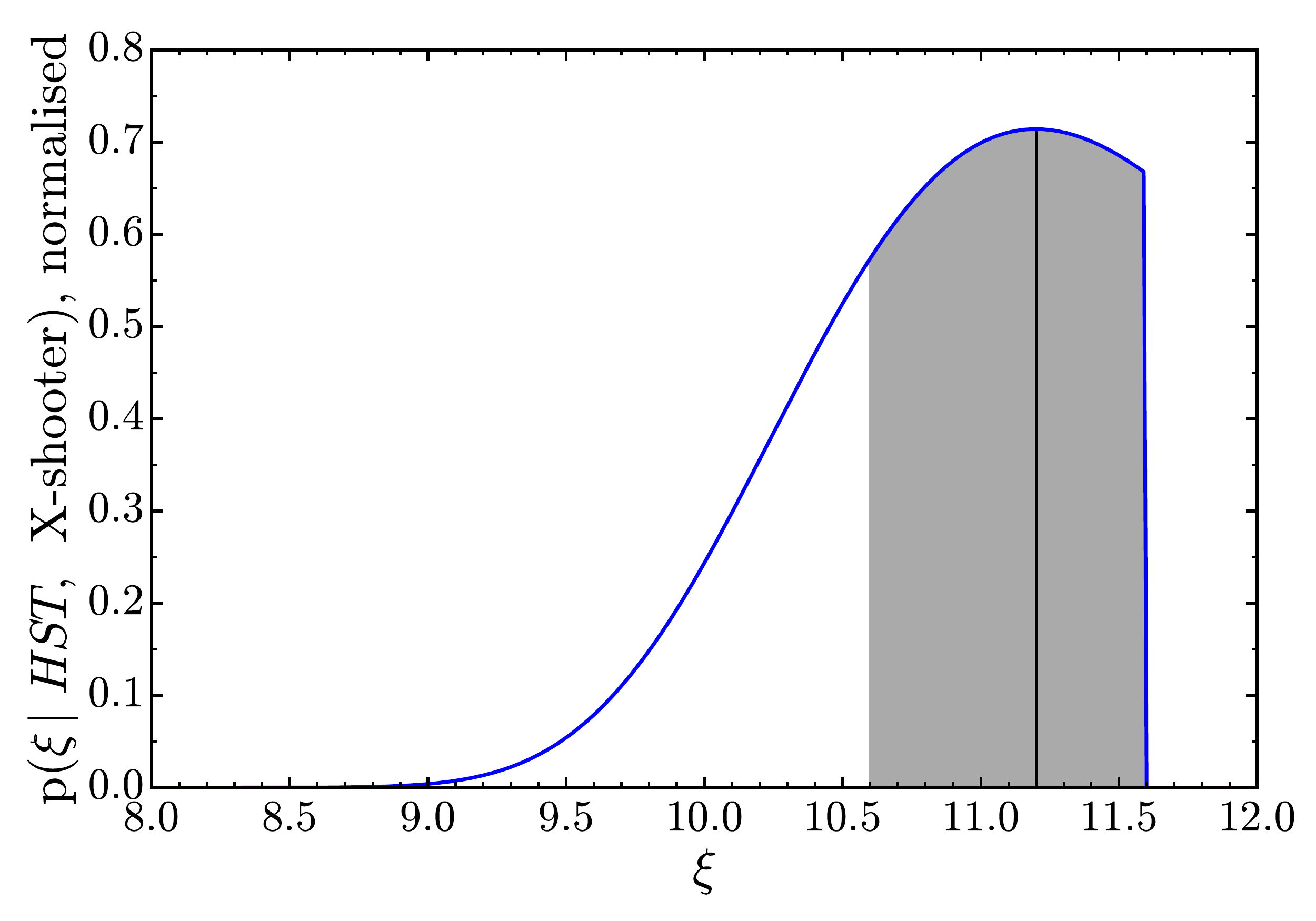}
\caption{Posterior distribution for \(\xi\), constrained by photometry
  and spectroscopy. The 68\% highest posterior density credible interval is shaded and the peak of the distribution is indicated at \(\xi = 11.2\). The cut-off at \(\xi = 11.6\) arises from determining a lower limit to the \(i_{814}\) flux from the transmission spectrum
derived in Sect.~\ref{sec:xs}.}
\label{pstr}
\end{figure}

The intrinsic quasar emission below \lya is modelled using the power
law chosen by \citet{Mortlock2011},~\(f_\lambda^{\,\rm intrinsic}
\propto \lambda^{-0.5}\), which is normalised to the observed spectrum
at 1280\,\AA. Absorption in the IGM is modelled using
Equation~\ref{eq:tauz}. The wavelength of each pixel is converted to a
\lya redshift, which is used to calculate \teff for a particular value
of \(\xi\). Measuring the absorbed spectrum with \textsc{synphot}
produces a synthetic \(i_{\rm814}\) magnitude that we can compare
directly to the F814W result. We show the fluxes measured for a range
of \(\xi\) values in Fig.~\ref{xis}. Combining the synthetic
photometry with spectroscopy, we determine a posterior distribution
for \(\xi\), shown in Fig.~\ref{pstr}. We have assumed a uniform
prior for \(\xi > 0\), and a Gaussian distribution for the
\(i_{\rm814}\) measurement which is equivalent to \((1.047 \pm 0.356)
\times 10^{-8}\)\,Jy. The sharp cut-off results from determining a
lower limit to the \(i_{\rm814}\) flux from the transmission spectrum
derived in Sect.~\ref{sec:xs}, measured as \(0.914 \times 10^{-8}\)\,Jy (AB
29.00, \(\textrm{S/N} \sim 30\)). Higher values of \(\xi\) are excluded, as the resulting \(i_{\rm814}\) measurements are below this secure lower flux limit. The posterior peaks at \(\xi = 11.2\), for which the synthetic
photometry yields the \emph{HST} result (Fig.~\ref{pstr}). The 68\% highest posterior density credible
interval is shaded, giving \(\xi =
11.2^{+0.4}_{-0.6}\). The resulting evolution of \teff\!\!\(\left(z\right)\) is shown in Fig.~\ref{tauevo}.

%SPIKE DISTRIBUTION

\subsection{Flux distribution in the \lya forest: the longest GP trough to date}
\label{sec:spikedist}

\begin{figure}[t]
\centering
\advance\leftskip-3cm
\advance\rightskip-3cm
\includegraphics[scale = 0.35, trim = 0cm 0.55cm 0cm 0.0cm, clip]{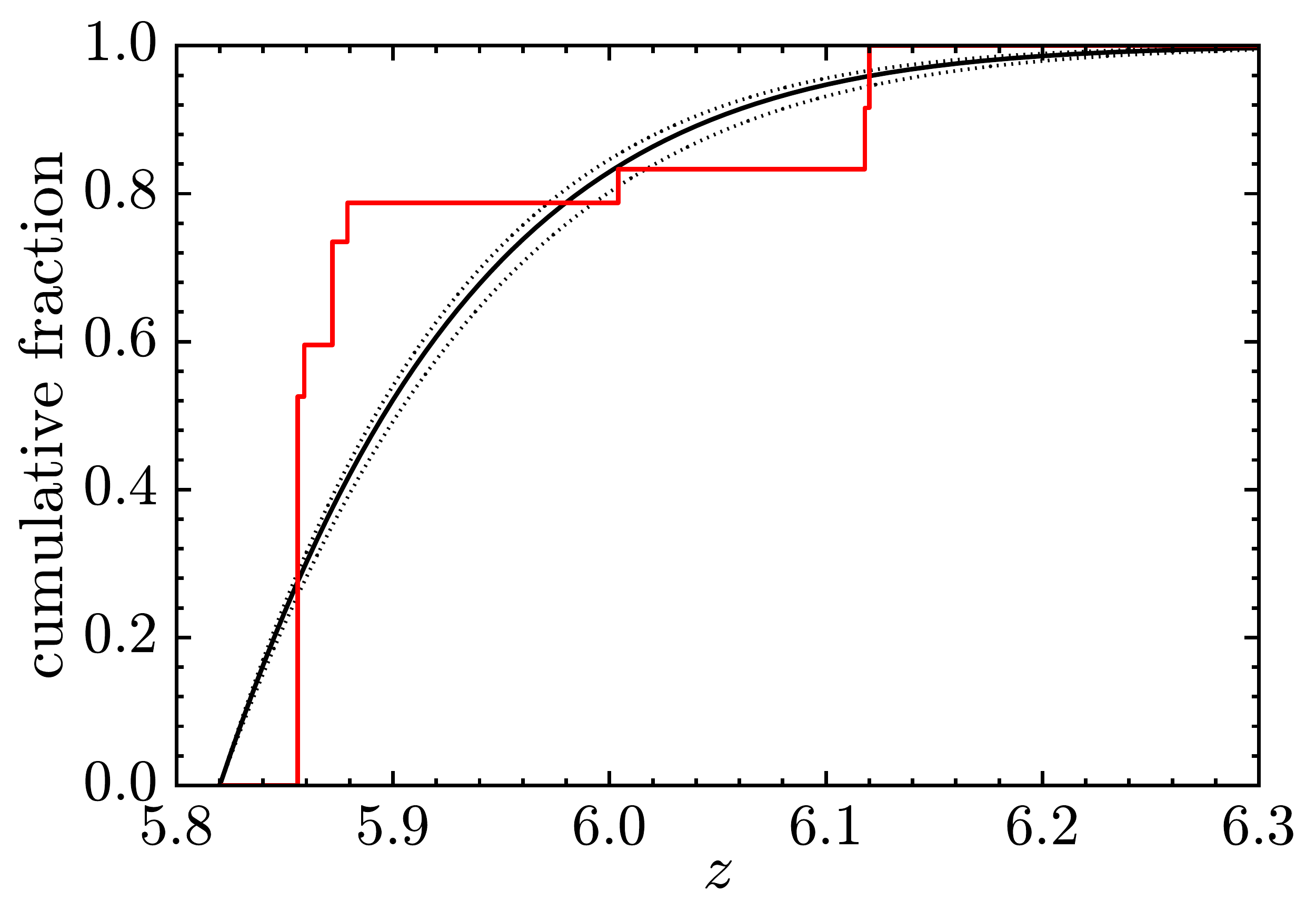}
\caption{Cumulative distribution of \lya forest flux contained in the
  detected transmission spikes (red), compared with the expected
  distribution of flux transmission based on our model for optical
  depth with \(\xi = 11.2\) (black; Sect.~\ref{sec:tau}). The dashed black lines indicate the uncertainty on our model for \teff.}
\label{cflux}
\end{figure}

The detected transmission spikes account for \((87 \pm 34\))\% of the
quasar flux detected by \emph{HST}, with the uncertainty dominated by
the photometry measurement. Therefore all of the transmission in the
spectrum may have been detected. Even if there are any additional
lower-significance transmission features they are likely to be
found in the same redshift range and to follow a similar
distribution. The shape of the cumulative distribution of the transmitted flux
therefore provides a noisy estimate of the shape of the cumulative transmission of
the \lya forest, and so it is interesting to compare this to the model. 

In Fig.~\ref{cflux} we present the cumulative proportion of flux
contained in transmission spikes with redshift i.e. the cumulative
transmitted flux as a function of redshift, normalised to the total
transmitted flux. The result from spectroscopy is compared to the
\teff model derived in Sect.~\ref{sec:tau}, which we use to produce an
average transmission spectrum. In both cases the distributions are
normalised to the total transmitted flux as we
wish to compare the relative shapes.

Broadly speaking, the distributions are in good agreement. In
particular, our model for \teff predicts effectively zero transmission
for \(z \gtrsim 6.2\), which is consistent with the transmission
spectrum. 

The distribution of transmission features is consistent with an extremely long
GP trough extending to the quasar's near-zone, covering the
redshift range \(6.12 \leq z \leq 7.04\), equivalent to a comoving
distance of 240\,\(h^{-1}\)Mpc. This is more than twice the length of the next
longest trough reported by \citet{Becker2015} towards the \(z = 5.98\)
quasar ULAS J0148+0600, although that trough extends to significantly
lower redshifts \((5.52 \leq z \leq 5.88)\).

%\citet{Fan2006} suggested that long Gunn-Peterson troughs will be able to provide useful constraints on the neutral hydrogen fraction at the highest redshifts. We therefore wish to find out what we can learn about reionization from the trough towards \uj.
%GB simulation work can go here

\subsection{Constraints on the IGM neutral fraction}

Previous studies have used the evolution of the IGM \lya opacity to
constrain $\xHi$ up to $z = 6.3$ \citep{Fan2006,Becker2015}. The large
cross-section of \lya means that the forest will be almost completely
absorbed by $z \sim 6$ even for $\xHi \sim 10^{-4}$
\citep[e.g.][]{Fan2006}, and density evolution will naturally make the
IGM even more opaque at $z > 6$.  Nevertheless, we wish to determine
what we can learn about $\xHi$ from the long trough towards \uj.

In order to constrain $\xHi$ we use artificial spectra generated from
the Sherwood suite of hydrodynamical simulations, which are described
in detail in \citet{Bolton2017}.  Briefly, they were run using a
modified version of the parallel Tree-PM SPH code P-GADGET-3, which is
an updated version of the publicly available GADGET-2
\citep{Springel2005}.  The models use a $\Lambda$CDM cosmology
consistent with recent {\it Planck} results \citep{Planck2014}.  Here
we use the reference 40\,-\,2048 model, which uses a 40\,$h^{-1}$ Mpc
box and $2 \times 2048^3$ particles, giving a gas particle mass of
$9.97 \times 10^{4}\,{\rm M}_\odot$.  Reionization occurs at $z_{\rm
  r} = 15$, whereafter the gas temperature and ionization fraction
evolve assuming a spatially uniform \citet{HM2012} ionizing
ultraviolet background (UVB).  Simulation outputs were captured in
redshift steps of $\Delta z = 0.1$, and one-dimensional \lya forest
skewers were generated from the outputs following standard methods
\citep[e.g.,][]{Theuns1998}.

Our goal is to set limits on the neutral fraction required to produce the 240\,$h^{-1}$ Mpc trough over $6.12 \le z \le 7.04$ seen towards \uj, and we generate mock \lya forest spectra with properties matched to the data using the following approach:

\begin{enumerate}
\item For each trial, six individual 40\,$h^{-1}$ Mpc skewers are
  selected at random from the simulation, with the simulation
  redshifts chosen to most closely match the observed line-of-sight.
\item The \lya optical depths (\(\tau_\alpha\)) are scaled according
  to the ratio of our desired $\xHi$ to the native $\xHi$ in the
  simulation.
\item Transmitted fluxes are computed as $F = e^{-\tau_{\alpha}}$. 
\item The spectra are smoothed to the nominal instrumental resolution
  of ${\rm FWHM} = 34\,{\rm km\,s^{-1}}$ and binned using
  10\,km\,s$^{-1}$ pixels, as in the real data.
\item Random Gaussian noise is added with dispersion given by the \uj
  error spectrum (normalized by a power law fit to the flux
  redward of the \lya emission line and extrapolated over the forest).
\item A matched-filter search for transmission peaks is then performed
  following the procedure described in Sect.~\ref{sec:spikes}.  For
  efficiency, we use only the narrowest filter width ($\Sigma =
  15\,{\rm km\,s^{-1}}$), although our results are not highly
  sensitive to this choice.

\end{enumerate}

%  For each trial, six individual 40\,$h^{-1}$ Mpc skewers are selected at random from the simulation, with the simulation redshifts chosen to most closely match the redshift evolution along the observed line-of-sight.  The \lya optical depths (\(\tau_alpha\)) are then scaled according to the ratio of our desired $\xHi$ to the native $\xHi$ in the simulation.  Transmitted fluxes are computed as $F = e^{-\tau_{\alpha}}$.  The spectra are then smoothed to the nominal instrumental resolution of ${\rm FWHM} = 34\,{\rm km\,s^{-1}}$ and binned using 10\,km\,s$^{-1}$ pixels, as in the real data.  Random Gaussian noise is added with an r.m.s. amplitude given by the \uj error array, which we normalize by a power law fit to the flux redward of the \lya emission line and extrapolated over the forest.  A matched filter search for transmission peaks is then performed following the procedure described in Section~\ref{sec:spikes}.  For efficiency, we use only the narrowest filter width ($\Sigma = 15\,{\rm km\,s^{-1}}$), although our results are not highly sensitive to this choice.

We parametrize the evolution in the neutral fraction as a power law of
the form $\xHi(z) = x_{0} [(1+z)/7.1]^\beta$, where $x_0$ is the
volume-weighted neutral fraction at $z = 6.1$, i.e., near the start of
the trough.  For reference, the simulation has a native $\xHi$
evolution given by $\xHi(z) \simeq (1.4 \times 10^{-4})
[(1+z)/7.1]^{9.1}$ over the redshift range of interest.  We generate
$10^4$ mock spectra for each combination of $x_0$ and $\beta$.  Our 95\% confidence limits on these parameters are those for which no
transmission peaks are detected with ${\rm S/N} \ge 5$ in at least 5\% of the trials.  For $\beta = 0$, we find $x_0 \ge 1.4 \times
10^{-4}$.  Increasing $\beta$ to 9 gives $x_0 \ge 1.1 \times 10^{-4}$,
which is consistent with using a \citet{HM2012} UVB.  The limits on
$x_0$ are thus largely insensitive to $\beta$, as expected if most of
the constraining power comes from the low-redshift end of the trough.

In summary, we find that the \lya trough towards \uj, while remarkably
long, provides only a weak lower limit to the volume-weighted neutral
fraction of the IGM ($\xHi \gtrsim 10^{-4}$), consistent with
a highly ionized IGM. The weakness of this constraint is because the trough lies at very high redshift ($z \ge 6.1$),
where large \lya opacities are more easily produced due to high IGM
densities.  We note that we have only considered an IGM model with a spatially uniform
UVB and temperature-density relation, which implicitly assumes that the most transmissive regions of
the IGM will be low-density voids. There is evidence that $\xHi$ is
significantly patchy on large scales at $z \gtrsim 5.5$ \citep{Becker2015}. Models that attribute this patchiness to a non-uniform background produced by
star-forming galaxies tend to predict that the UVB intensity will be
anti-correlated with large-scale density enhancements
\citep{Davies2016,DAloisio2016}.  In these models, the highest transmission tends to come from
high-density regions, and thus a lower $\xHi$
would be required to produce significant \lya transmission.
Alternatively, the $\xHi$ patchiness may be due to residual large-scale temperature fluctuations from an extended reionization process \citep{DAloisio2015}.  In that model, high transmission tends to occur in hot ($T \sim 20,000\,\textrm{-}\,30,000~{\rm K}$), low-density regions that have only recently been reionized. Since these regions would be hotter than we have assumed and the neutral fraction scales as $T^{-0.7}$, \lya transmission might appear even if the globally-averaged $\xHi$ is somewhat larger than our constraint.

The constraints on \(\xHi\) from the \lya trough towards \uj
  are considerably weaker than those determined from \lya damping wing
  absorption, first measured by \citet{Mortlock2011}, who found \(\xHi
  > 0.1\) at \(z = 7.1\). The most recent \lya damping wing
  constraints on the volume-weighted neutral fraction come from an
  analysis of the FIRE infrared spectrometer spectrum of \uj
  \citep{Simcoe2012} by \citet{Greig2016}, who find \(\xHi\,(z =
  7.1) = 0.4 \pm 0.2 \). \citet{Greig2016b} argue that the measured
  \lya damping wing absorption towards \uj provides one of the strongest constraints on reionization that are currently available. As we begin to probe ever-higher neutral fractions, measurements of \lya damping wing absorption will likely provide far more useful constraints than the \lya forest. A further analysis of the damping wing towards \uj using the deeper X-shooter spectrum will be presented in Hewett et al. (\emph{in prep}.). 

%Our lower
%limits on $\xHi$ derived assuming a uniform UVB are therefore likely
%to be robust.

\section{Summary}
\label{sec:sum}

%Combining \emph{HST} F814W photometry with a new X-shooter spectrum,
%we have presented an analysis of the \lya forest towards the
%\(z=7.084\) quasar \uj. Here we summarise the main results of the
%paper:
We have analysed a deep X-shooter spectrum and {\it HST} F814W photometry of \uj to probe the \lya forest to \(z \simeq 7\) along the line of sight to the quasar. Our main results are as follows: 

\begin{enumerate}

\item We revise F814W photometry of \uj to be \(i_{\rm814} = 28.85\),
  with \({\rm S/N} = 2.9\). For a faint source such as \uj which is measured
  against a high background, the \citet{Anderson2010} CTE correction
  included in the ACS pipeline merely introduces noise to the
  photometry.

\item Significant transmission in the Lyman series forest is
  restricted to the redshift range $5.856<z<6.121$, where seven spikes
  were detected at ${\rm S/N}>5$.

\item Photometry provides us with more useful constraints on effective
  optical depth, while spectroscopy can still be used to derive a
  lower limit on transmitted flux. Using data for \uj alone and
  assuming a fixed normalization at \(z = 5.5\), for redshifts \(z >
  5.5\) we find \teff\(\propto (1+z)^{\xi}\) where \(\xi =
  11.2^{+0.4}_{-0.6}\).

\item  We detect an
  extremely long dark gap extending from the final transmission spike
  to the quasar's near-zone, equivalent to a comoving distance of
  240\(\,h^{-1}\)Mpc. The GP trough nevertheless provides only a weak
  limit on the IGM neutral fraction
  $\xHi \gtrsim 10^{-4}$ at $95\%$ confidence. This is a generic
  limitation of observations of the \lya forest at such high redshifts.

\end{enumerate}

\begin{acknowledgements}

We are grateful to the referee, Michael Strauss, who provided constructive comments and improvements. We would like to thank the ACS instrument team for their helpful
guidance on the effects of CTE losses. We also thank Jamie Bolton for
the use of the Sherwood simulations. This work was supported by Grants ST/K502042/1 and ST/N000838/1 from the Science and Technology Facilities Council. SJW gratefully acknowledges the
support of the Leverhulme Trust through the award of a Leverhulme
Research Fellowship. GDB is supported by the National Science
Foundation under Grant AST 1615814. PCH and RGM acknowledge support
from the STFC via a Consolidated Grant to the Institute of Astronomy, Cambridge. BPV acknowledges funding through the ERC grant ``Cosmic Dawn''. \\

\end{acknowledgements}

%\bibliography{lyapaperrefs}{} \bibliographystyle{aa}
\end{document}